\documentclass[a4paper,11pt]{article}
\pdfoutput=1 

\usepackage{jcappub} 

\usepackage[T1]{fontenc} 

\usepackage{amsfonts}
\usepackage{amsmath}
\usepackage{amssymb}
\usepackage{graphicx,color}
\usepackage{float}
\usepackage{hyperref}
\usepackage{subfigure}
\usepackage{dcolumn}
\usepackage{soul}
\usepackage{verbatim}
\usepackage[normalem]{ulem}
\usepackage{bm}  
\usepackage{booktabs}
\title{Sourcing vector perturbations in warm inflation through shear viscous effects}
\author[a]{Arjun Berera,}
\author[a,b]{Suddhasattwa Brahma,}
\author[a,1]{Zizang Qiu,\note{Corresponding author. \href{mailto:Zizang.Qiu@ed.ac.uk}{\texttt{Zizang.Qiu@ed.ac.uk}}}}
\author[c]{Rudnei O. Ramos}

\abstract{Vector modes are generally believed not to play a significant role in standard cosmological perturbation theory because they decay rapidly. Non-equilibrium dynamics in the radiation bath present during warm inflation can lead to shear viscous effects, requiring a description beyond the perfect-fluid approximation. We adopt the fluctuating hydrodynamic approach to account for stochastic anisotropic stress related to shear viscosity. We show that the vector mode obeys a constrained Langevin equation in an inflationary background. The resulting power spectrum, considered in the weak-viscosity regime, is proportional to the shear viscosity and it also depends on the Hubble slow-roll parameter $\epsilon$, decaying approximately as $a^{-(1-\epsilon)}$. This behaviour is significantly slower than the unsourced, non-viscous power spectrum, which decays as $a^{-4}$. The continuously sourced vector modes, nevertheless, decay on superhorizon scales, and the illustrative benchmark parameter choices adopted in our numerical examples give a negligible late-time vector-to-scalar ratio. Despite the absence of direct large-scale observable vector signatures, the main results establish a source of vector perturbations during warm inflation and provide a framework for studying possible secondary effects along with large-scale white noises in the vector channel.}

\affiliation[a]{Higgs Centre for Theoretical Physics, School of Physics and Astronomy, University of Edinburgh, Edinburgh EH9 3FD, UK}
\affiliation[b]{Physics and Applied Mathematics Unit, Indian Statistical Institute, 203 B.T. Road, Kolkata 700108, India}

\affiliation[c]{Departamento de F\'{\i}sica Te\'orica, Universidade do Estado do Rio de Janeiro, 20550-013 Rio de Janeiro, RJ, Brazil }

\emailAdd{ab@ed.ac.uk}
\emailAdd{suddhasattwa.brahma@gmail.com}
\emailAdd{Zizang.Qiu@ed.ac.uk}
\emailAdd{rudnei@uerj.br}

\begin{document}
\maketitle

\section{Introduction}
\noindent 
Vector perturbations are generally believed not to play an important role in cosmology since they decay rapidly, unlike the usual adiabatic scalar and tensor modes, which can approach constant values on superhorizon scales~\cite{Weinberg2008,Baumann2022,MalikWands2009}. Inhomogeneous vector metric perturbations can be generated by relative fluid motion and local shear deformation geometrically. Given a non-vanishing shear viscosity coefficient of the fluid, this gives rise to anisotropic shear stress, the presence of which leads to non-trivial vector mode dynamics and changes its characteristic redshifting. Indeed, realistic dissipative processes present in the universe necessitate deviation from a perfect fluid approximation that is often adopted at the background level, and the inclusion of bulk and shear viscosities, which are local transport coefficients that encode internal friction when different neighbouring fluid particles move with different velocities. 

For a perfect fluid, the stress-energy tensor is
\begin{equation}
T_{\mu\nu}^{\rm pf}=(\rho+p)u_\mu u_\nu-pg_{\mu\nu}\,,
\label{eq:perfect-fluid-stress}
\end{equation}
for a mostly-minus metric signature $(+,-,-,-)$. Here $\rho$ and $p$ are the fluid rest-frame energy density and pressure. $u_{\mu}$ is the 4-velocity of the fluid rest-frame in a cosmological Friedmann-Lema\^itre-Robertson-Walker (FLRW) background which satisfies $u^\mu u_\mu=1$ and uniquely selects a time-like direction. We use the Landau-Lifshitz frame, in which the local rest frame is defined by the absence of energy flux, $T_{\mu\nu}u^\nu=\rho u_\mu$. An alternative but physically equivalent frame choice results in the Eckart formulation of relativistic hydrodynamics, in which the rest fluid frame is chosen to be the frame of charge flow~\cite{Romatschke2010,Kovtun2012}. Orthogonal to the time-like direction $u^{\mu}$, the spatial metric $h_{\mu\nu}$ defined by~\cite{TsagasChallinorMaartens2008}:
\begin{equation}
    h_{{\mu\nu}}=g_{\mu\nu}-u_{\mu}u_{\nu}.
\end{equation}
can serve as a spatial projector. Assuming that the velocity gradient is small, let us follow the standard first-order Landau-Lifshitz formulation of  relativistic viscous hydrodynamics. The viscous part $\Pi_{\mu\nu}$ of the total stress energy tensor $T_{\mu\nu}=T_{\mu\nu}^{\mathrm{\rm pf}}+\Pi_{\mu\nu}$
is given by~\cite{Landau2004}
\begin{equation}
    \Pi_{\mu\nu}= \zeta_s(h_{\mu\alpha}\nabla^{\alpha}u_{\nu}+h_{\alpha\nu}\nabla^{\alpha}u_{\mu}-\frac{2}{3}h_{\mu\nu}\nabla^{\alpha}u_{\alpha})+\zeta_b h_{\mu\nu}\nabla^{\alpha}u_{\alpha},
    \label{eq_bulk}
\end{equation}
which explicitly satisfies $
\Pi_{\mu\nu}u^\nu=0$ due to spatial projection. Here $\nabla^{\alpha}$ denotes the covariant derivative associated with $g_{\mu\nu}$. In Eq.~\eqref{eq_bulk}, $\zeta_s>0$ and $\zeta_b>0$ are the shear and bulk viscosity coefficients, respectively. These viscosity coefficients in general depend on temperature and pressure. In a radiation bath with vanishing chemical potential, these coefficients are functions of temperature alone. The first-order constitutive relation above assumes that the viscous stress adjusts instantaneously to reflect changes in local velocities, which nonetheless approximates the low-frequency, long-wavelength limit of the causally completed hydrodynamical theory well. The validity of this first order relativistic formalism being applied in an expanding spacetime will be discussed in more detail in Section~\ref{subsec:asymptotic-consistency}. Defining the expansion scalar $\theta$ and the shear itself $\sigma_{\mu\nu}$ as
\begin{equation}
\theta=\nabla_\mu u^\mu,\qquad
\sigma_{\mu\nu}=h_\mu{}^\alpha h_\nu{}^\beta\left(\nabla_{(\alpha}u_{\beta)}-\frac{1}{3}h_{\alpha\beta}\theta\right),
\label{eq:expansion-shear-definitions}
\end{equation}
where the parentheses indicate symmetrisation, the viscous stress can be recast as
\begin{equation}
\Pi_{\mu\nu}=2\zeta_s\sigma_{\mu\nu}+\zeta_b h_{\mu\nu}\theta.
\label{eq:viscous-stress-compact}
\end{equation}
In a local instantaneously comoving frame, where $u_i=0$ and $u^0=u_0=1$, the non-vanishing components of the manifestly covariant dissipative stress tensor above read
\begin{equation}
\Pi_{ij}=2\zeta_s\sigma_{ij}-\zeta_b\delta_{ij}\theta\,.
\label{eq:viscous-stress-local-spatial}
\end{equation}

Viscous processes are irreversible transfer of momentum between neighbouring fluid elements with different velocities. Bulk viscosity contributes to the isotropic background pressure and has been found to be connected to dark sectors in other cosmological settings~\cite{Colistete2007}. Unlike bulk viscosity, the deterministic contribution of its shear counterpart vanishes on an exactly homogeneous and isotropic background. However, shear viscous effects enter at the level of perturbation and become coupled to vector modes of the metric perturbation. 

Warm inflation provides a natural dynamical setting in which shear viscosity can be generated and maintained during early cosmic expansion as a consequence of small deviations from equilibrium arising from intrinsic microphysical couplings of the bath dynamics~\cite{BasteroGil:2011Shear,BasteroGil:2014}. In the quasi-stationary regime of warm inflation, the viscous coefficients can be approximately constant given that the physical temperature of the universe evolves slowly on cosmic timescales. Interactions continuously replenish a radiation bath, while internal bath interactions maintain approximate local equilibrium. Shear viscous effects in warm inflation have been studied in connection with scalar cosmological perturbations~\cite{BasteroGil:2011Shear,BasteroGil:2014}. This work will make the extension to vector perturbations by building on a key observation, that is, according to the Landau-Lifshitz formalism of fluctuating hydrodynamics, the deterministic viscous stress is always accompanied by a zero-mean stochastic stress whose two-point correlator is statistically fixed by the local fluctuation-dissipation relation~\cite{Landau2004,BasteroGil:2014,Landau1957} and has a non-trivial transverse contribution. Schematically, deterministic shear relaxes vector perturbations, whilst its stochastic counterpart, the random anisotropic stress, continuously sources them. Fluctuating hydrodynamical evolution thus goes beyond the frequently encountered mean-field approximation in cosmology and provides important insights into how short-scale physics affects large-scale physics. A causal hydrodynamic reformulation such as M\"uller--Israel--Stewart theory can also be used to determine the corresponding finite-relaxation-time fluctuations~\cite{KumarBhattMishra2014}.

The rest of the paper is organised as follows. In section~\ref{sec:vector-dynamics}, we set up the standard vector mode
dynamics in an FLRW spacetime background. We derive the momentum constraint and sourced Langevin equation for the vector perturbation, taking into account both shear damping effects and stochastic noises arising from the warm-inflationary cosmological background. In section~\ref{sec:sourced-spectrum}, using the fluctuation-dissipation relation, we can fix the
two-point noise correlator and obtain the vector power spectrum
by computing the projected noise kernel. Section~\ref{sec:asymptotic-regimes} studies the late-time asymptotics of the power spectrum integral and distinguishes the physically accessible Hubble-dominated regime from the formal viscosity-dominated regime. Consistency conditions from both the causal completion of the Landau-Lifshitz hydrodynamics with a finite shear relaxation time and the warm inflationary background are invoked to elucidate their domains of validity. In Section~\ref{sec:discussion}, we explain how the decay of the sourced vector mode is significantly slowed down in comparison
to the free evolution and connect the presence of superhorizon stochastic noise in warm inflation to recently emphasised large-scale white noise~\cite{BarenboimIrelandStebbins2026PRL}. We conclude in section~6. Three appendices are also included where we give complementary technical details of relevant computations.

We use natural units $\hbar = c = k_B = 1$, while keeping the reduced Planck mass $M_{\rm Pl}$ explicit through $\kappa = 8\pi G = M_{\rm Pl}^{-2}$ as a convention.

\section{Vector Mode Dynamics}
\label{sec:vector-dynamics}
In the spatially flat FLRW spacetime, the background metric is given by
\begin{equation}
g_{\mu\nu}^{(0)}\,dx^\mu dx^\nu
= a^2(\eta)\left(d\eta^2-\delta_{ij}\,dx^i dx^j\right),
\end{equation}
where $a(\eta)$ is the scale factor as a function of the conformal time. In this paper, dots and primes denote derivatives with respect to cosmic time $t$ and conformal time $\eta$, respectively, with $dt = a\,d\eta$. Throughout, we distinguish the physical Hubble rate $H$ from the conformal Hubble rate ${\cal H}$, where
\begin{equation}
    H \equiv \frac{\dot a}{a},\qquad
    \mathcal{H} \equiv \frac{a'}{a} = aH,\qquad
    \epsilon \equiv -\frac{\dot H}{H^2},
\end{equation}
with $\epsilon$ the Hubble slow-roll parameter. Perturbations around this background metric can be written most generally as
\begin{equation}
   ds^2=a^2(\eta)[(1+2 A)d\eta^2+2B_idx^id\eta-(\delta_{ij}-2F_{ij})dx^idx^j].
   \label{eq:SVT}
\end{equation}
In the synchronous gauge, $A=B_i=0$ and keeping only the vector part of the spatial perturbation, we have that
\begin{equation}
    \Tilde{g}_{ij}=a^2(\partial_jF_i+\partial_iF_j),\qquad \partial_i F^i=0.
\end{equation}
In the equation above, the vector mode $F_i = F_i(\eta, \mathbf{x})$ is chosen to be divergenceless following the standard Scalar-Vector-Tensor (SVT) decomposition of metric perturbations~\cite{Baumann2022}. Hence, we find
\begin{eqnarray}
    ds^2 = a^2(\eta) \left[d\eta^2 - \left(\delta_{ij} - \partial_i F_j - \partial_j F_i \right) dx^idx^j\right]\,.
    \label{eq:vector-metric-line-element}
\end{eqnarray}
\subsection{Linearised Einstein's equations}
For the metric in Eq.~\eqref{eq:vector-metric-line-element}, the nonzero components of the Einstein tensor can be obtained up to first order:
\begin{align}
    &G^0_0=\frac{3a'^2}{a^4},\\
    &G^0_i=-\frac{\nabla^2F_i'}{2a^2},\label{eq:einstein-tensor-momentum}\\
    &G^i_j=-\left(\frac{a'^2}{a^4}-\frac{2a''}{a^3}\right)\delta^i_j+\left(\frac{\partial^iF''_j}{2a^2}+\frac{\partial_jF''^i}{2a^2}+\frac{a'\partial^iF'_j}{a^3}+\frac{a'\partial_jF'^i}{a^3}\right),
\end{align}
where here $\nabla^2 \equiv \delta^{ij}\partial_i\partial_j$ is the comoving coordinate Laplacian. The relation to the physical spatial Laplacian is given in Appendix~\ref{app:lif-flrw}.

The metric perturbation has to be accounted for, according to the Einstein equation, by the corresponding velocity perturbation in the stress energy tensor of the matter distribution. The matter content for a toy model of warm inflation that we are going to consider here consists of a scalar inflaton interacting with a radiation fluid. Let $\rho$ and $p$ denote the total background energy density and pressure of a cosmological medium. Contributions from the inflaton field $\phi$ are
\begin{equation}
\rho_\phi=\frac{\phi'^2}{2a^2}+V,\qquad
p_\phi=\frac{\phi'^2}{2a^2}-V,
\end{equation}
where $V$ is the inflaton primordial potential. We distinguish the total enthalpy density $w$ from the radiation enthalpy density $w_r$:
\begin{equation}
    w \equiv \rho + p = \dot{\phi}^{2} + w_r,\qquad
    w_r \equiv \rho_r + p_r = \frac{4}{3}\rho_r.
    \label{eq:enthalpy-dictionary}
\end{equation}
A slow-rolling inflaton dominates the background energy and pressure, but as a scalar field its linear momentum perturbation carries no transverse part~\cite{MalikWands2009}, and hence vector perturbations must be sourced solely by the radiation fluid.

Let us define the perturbed velocity of the radiation fluid by $u_r^\mu=a^{-1}(1,v_r^i)$. The viscous stress must be evaluated using the radiation velocity, since it is the radiation field that carries the transport coefficient~\cite{BasteroGil:2014}. Given these conditions, the various components of perturbed stress-energy tensor in the fluid rest frame are
\begin{align}
&T^0{}_0=\rho\,,\\
&T^0{}_i=-w_r v_{r i}\,,\\
&T^i{}_j=-p\delta^i{}_j + \Pi^i_j\,,
\label{eq:stress-radiation}
\end{align}
where $v_{ri}$ is the peculiar velocity perturbation with its spatial index lowered with $\delta_{ij}$, and $\Pi^i_j$ is the viscous shear stress. In comoving coordinates, this shear stress is again given by
\begin{eqnarray}
\Pi^i_j = \frac{\zeta_s}{a}
\left(\partial^iv_{r j}+\partial_jv_r^i-\partial^iF'_j-\partial_jF^{i\prime}\right)\,.
\label{eq:shear_stress}
\end{eqnarray}
The terms involving $F'$ arise because the shear is defined using the perturbed spacetime geometry. By writing down Eq.~\eqref{eq:shear_stress}, we have assumed that the anisotropic stress tensor relaxes instantaneously to local shear as mentioned and reaches its Navier-Stokes value $\Pi^i{}_{j,\mathrm{NS}}$ on the right-hand side of Eq.~\eqref{eq:shear_stress}. For related analysis on causality and linear stability in the finite relaxation time extension to Israel--Stewart theory, see~\cite{BemficaDisconziNoronha2019,SammetMayerRischke2023}. 

We will drop the subscript for the velocity perturbation and denote it as $v_i$ with its radiation origin implied. The first order Einstein's equations $G^\mu_\nu = \kappa T^\mu_\nu$ will yield the necessary constraint and evolution equations for the vector modes. The diagonal components yield the Friedmann and Raychaudhuri equations:
\begin{equation}
    3H^2 = \kappa\rho,\qquad
    \dot H = -\frac{\kappa}{2}w.
    \label{eq:raychaudhuri}
\end{equation}
Consequently,
\begin{equation}
    \kappa w = 2\epsilon H^2,\qquad
    \kappa a^2 w = 2\epsilon\mathcal{H}^2.
\end{equation}
The transverse part of the $(0i)$ component will give the so-called diffeomorphism or momentum constraint\footnote{In the Hamiltonian formulation of General Relativity, diffeomorphism invariance imposes a series of constraint equations in the total phase space. This can be seen most clearly in the generally covariant ADM formulation~\cite{ADM1962}. Here, this equation relating $F_i'$ and $v_i^V$ is a spatial diffeomorphism constraint obtained by varying the ADM shift $N^i$.}:
\begin{eqnarray}
    \frac{\nabla^2F'_i}{2a^2} = \kappa w_r v_i^{V}=\kappa w_r\mathsf P_i{}^jv_j\,,
\label{eq:einstein-momentum}
\end{eqnarray}
where the transverse projector is $\mathsf P_i{}^{j}=\delta_i{}^{j}-\frac{k_i k^{j}}{k^2}$. The vector metric perturbation and transverse fluid velocity are therefore not independent initial gravitational data. As it will be evident from below, only the transverse component of the velocity perturbation participates in vector mode dynamics.

Linear perturbations can always be projected onto the basis of Laplacian eigenfunctions as 
\begin{equation}
F_i(\eta,\mathbf{x})=F(\eta)Q_i(\mathbf{x})\,,
\qquad
 v_i(\eta,\mathbf{x})=v^{V}(\eta)Q_i(\mathbf{x})\,.
\label{eq:harmonic-expansion}
\end{equation}
The spatial Laplacian eigenfunctions \(Q_i\) are chosen to be divergenceless and satisfy 
\begin{equation}
\nabla^2 Q_i^{(\lambda)}=-k^2Q_i^{(\lambda)},
\qquad
\partial^iQ_i^{(\lambda)}=0\, ,
\label{eq:vector-harmonics}
\end{equation}
where \(\nabla^2\equiv\delta^{ij}\partial_i\partial_j\) is again the Laplacian of the Euclidean 3-space and \(k\) is the usual comoving wavenumber. For each $k$, the eigenfunction condition together with the transversality condition leaves exactly two linearly independent vector harmonics $Q_i^{(\lambda)}(\mathbf x)$ labelled by transverse polarizations $\lambda=1,2$. The transverse unit polarization vectors $e_i^{(\lambda)}(\hat{\mathbf k})$ in Fourier space satisfy
\begin{equation}
    k^i e_i^{(\lambda)}=0,\qquad
    e_i^{(\lambda)}e^{(\lambda')i}=\delta^{\lambda\lambda'},\qquad
    \sum_{\lambda=1}^{2}e_i^{(\lambda)}e^{(\lambda)j}=\mathsf P_i{}^{j}.
\end{equation}
In what follows, the explicit superscripts $(\lambda)$ and $V$ will be suppressed and the transverse scalar amplitudes such as $F(\eta)$ will be taken to apply to both polarizations. In the final result, we can account for the polarisation sum in the total vector power spectrum multiplying by a factor of $2$. This is because the thermal state of the radiation fluid is spatially isotropic and we will not consider any parity-violating interactions in this work, so the two polarization spectra are equal. 
We give more details of how the comoving and the full physical Laplacian(-Beltrami) operator are related in appendix~\ref{app:lif-flrw}.

For a fixed non-zero wave-vector and a given polarization, we can define new variables~\cite{Grishchuk1993,PintoNeto:2020}
\begin{equation}
h(\eta)=-kF(\eta)\,,\qquad \omega(\eta)=-a(\eta)^2w_r(\eta)v(\eta)\,.
\label{eq:reduced-variables}
\end{equation}
This further simplifies the constraint in Eq.~\eqref{eq:einstein-momentum} when we project it onto the same harmonic basis:
\begin{equation}
-\frac{k}{2}h'=\kappa\omega\,.
\label{eq:momentum-constraint}
\end{equation}
The traceless part of the $(ij)$-component of linearized Einstein's equations governs the dynamical evolution of the vector perturbation, leading to
\begin{eqnarray}
\frac{\partial^iF''_j+\partial_jF^{i\prime\prime}}{2a^2}
+\frac{a'}{a^3}\left(\partial^iF'_j+\partial_jF^{i\prime}\right)
 =\frac{\kappa\zeta_s}{a}
\left(\partial^iv_j+\partial_jv^i-\partial^iF'_j-\partial_jF^{i\prime}\right)\,.
\label{eq_ij}
\end{eqnarray}
From the above, we can see that the Friedmann equations depend on the total density and pressure, as expected, while the transverse viscous response is where the effective inertial mass density $w_r$, also referred to as the enthalpy density when emphasising the thermodynamic properties of the fluid, enters. Regrouping the shear viscosity coefficient into a new scalar variable~\cite{Grishchuk1993,PintoNeto:2020} 
\begin{equation}
    \chi(\eta)=\zeta_s(\eta)a(\eta) k^2(F'(\eta)-v(\eta)),
\end{equation}
we can use Eq.~\eqref{eq:momentum-constraint} and rewrite Eq.~\eqref{eq_ij} above as 
\begin{eqnarray}\label{eq:without_noise_vector}
    \omega'+2\frac{a'}{a} \omega  + \chi =0\,.
\end{eqnarray}
The physical gauge-invariant vector metric variable is the shear combination $\Sigma_i = B_i - F_i'$. Indeed, under a vector gauge transformation, $B_i \to B_i + \xi_i'$ and $F_i \to F_i + \xi_i$, so $\Sigma_i$ is invariant. In synchronous gauge,
\begin{equation}
    \Sigma_i = -F_i'\,.
    \label{eq:synchronous-vector-shear}
\end{equation}
Thus, $F_i'$ is the representative gauge-invariant vector metric shear, whereas a time-independent part of $F_i$ is a residual gauge freedom. Eq.~\eqref{eq:without_noise_vector} is first order in the reduced physical variable as a result of the momentum constraint. This is in direct contrast with the second order nature of Mukhanov-Sasaki equations for gauge invariant scalar and tensor perturbations~\cite{Weinberg2008,Baumann2022,MalikWands2009}. However, at this stage, we have simply assumed the presence of a transverse velocity perturbation and shear viscous effects in the fluid without discussing its physical origin. In order to do so, let us first describe the background inflationary evolution we have in mind.

\subsection{Warm background dynamics}
\label{sec:warm-background}
So far, we have been fairly agnostic regarding the background spacetime except for assuming it to be a cosmological one. To specify the background, we first need to set up our physical scenario. In warm inflation, expansion is driven by a scalar inflaton field $\phi$ while interactions between the inflaton and radiation create a fluctuation-dissipation dynamics and produce a dynamically sustained radiation bath during inflation~\cite{Berera1995,BereraMossRamos2009,Kamali:2023lzq,Bastero-Gil:2016qru,5BerghausDrewesZell2025,BereraFang1995}. The background warm inflation equations of motion for the inflaton and radiation fluids are
\begin{align}
    &3\mathcal H^2 = \kappa a^2\left(\rho_\phi+\rho_r\right)\,,\\
    &\ddot{\phi}+(3H+\Upsilon)\dot{\phi}+V_\phi = 0\,,\\
    &\rho_r'+4\mathcal H\rho_r = \frac{\Upsilon}{a}\phi'^2\,,
\end{align}
where the dissipation coefficient $\Upsilon$ denotes the energy transfer from the inflaton to the radiation bath. In the slow-roll regime, consider the quasi-stationary radiation bath with $ \rho_r'\sim0$ to be close to equilibrium given that thermal relaxation time $\tau_{\rm therm}$ is short compared to cosmic expansion $H\tau_{\rm therm}\ll1$~\cite{HallMossBerera2004,BereraMossRamos2009,BasteroGil2012Stability}. The background equations simplify to
\begin{equation}
    3H(1+Q)\dot{\phi} \simeq -V_\phi,\qquad
    4H\rho_r \simeq \Upsilon\dot{\phi}^{\,2},\qquad
    Q \equiv \frac{\Upsilon}{3H},
    \label{bg}
\end{equation}
where $Q$ is the dissipation ratio. These background quantities determine the time evolution of the scale factor $a(\eta)$ and the local temperature ${\mathcal T}(\eta)$ of the bath, which will enter the perturbation equations. In warm inflation, the radiation bath can be taken to be approximately quasi-stationary, \textit{i.e.,} the fractional rate $|\rho_r'|/(\mathcal H\rho_r)$ is small rather than being identically zero. With fixed relativistic degrees of freedom $g_*$, this makes $|\mathcal T'|/(\mathcal H\mathcal T)$ accordingly small \footnote{Note for a fixed effective number of relativistic degrees of freedom, under thermal equilibrium 
$\rho_r=\frac{\pi^2}{30}g_*\mathcal T^4$ and it follows that$\frac{\mathcal T'}{\mathcal H\mathcal T}
=\frac14\frac{\rho_r'}{\mathcal H\rho_r}$. Under the given approximation we can take the physical temperature to be approximately constant.}. In the following calculation, we will treat the physical temperature as effectively time-independent $\mathcal T(\eta)=\mathcal T_0$ as a further simplification of Eq.~\eqref{eq:without_noise_vector}.

Using the Raychaudhuri equation in Eq.~\eqref{eq:raychaudhuri}, we can express the total enthalpy, also known as inertial mass density $w=\rho+p$  in terms of the slow-roll parameter $\epsilon = - \dot{H}/H^2$ of the above quasi-de Sitter background
\begin{equation}
\rho+p=-\frac{2\dot H}{\kappa}\,,
\qquad
\kappa w=2\epsilon H^2\,.
\label{eq:slow-roll-enthalpy}
\end{equation}
From Eq.~\eqref{bg} and the definition of dissipation ratio, the total inertial mass density $w$ is related to the radiation inertial mass density $w_r$ as
\begin{equation}
w_r\simeq Q\dot{\phi}^2,
\qquad
\frac{w_r}{w}\simeq\frac{Q}{1+Q},
\end{equation}
which satisfies
\begin{equation}
    0<w_r\leq w = 2\epsilon M_{\rm Pl}^2H^2\,.
\end{equation}
For our purposes, we shall assume that the slow-roll parameter is a small constant. With the normalization $a(\eta_*)=1$ and $H_*=H(\eta_*)$ at reference time $\eta^*$, define
\begin{equation}
    \beta=\frac{1}{1-\epsilon},\qquad
    A_0=\left[\frac{1}{H_*(1-\epsilon)}\right]^\beta\,.
\end{equation}
The background evolution is
\begin{equation}
    a(\eta)=A_0(-\eta)^{-\beta},\qquad
    \mathcal H=-\frac{\beta}{\eta},\qquad
    H=H_*a^{-\epsilon}\,.
    \label{eq:constant-epsilon-background}
\end{equation} Under the quasi-equilibrium assumption, it is straightforward to see that $\frac{w_r}{w}$ increases with time given the slow-rolling inflaton background. Thus, the calculation for vector perturbations done in this work is only valid when the radiation energy is less than the total energy $w_r\leq w$. 

Under equilibrium, the fluid stress-energy tensor is given by that of the perfect fluid and the shear stress vanishes. Microscopic decay processes within the warm-inflation radiation bath could result in a small departure from equilibrium encoded by shear and bulk viscosities~\cite{BasteroGil:2011Shear,BasteroGil2012Stability}. When neighbouring fluid elements drift towards different directions relative to each other, this leads to internal fluid friction. Unlike momentum flow, this is associated with entropy production and thus thermodynamically irreversible. This is encoded by the shear viscous stress, the symmetric and traceless part of the velocity gradient in Eq.~\eqref{eq:shear_stress}, following the first-order Landau-Lifshitz formalism of hydrodynamics which applies for sufficiently short radiation bath relaxation time and shear relaxation time $\tau_{\rm \pi}$,
\begin{equation}
    H\tau_{\rm therm}\ll 1,\qquad H\tau_\pi\ll 1\,.
    \label{eq:bath-relaxation-conditions}
\end{equation}
Note this is compatible with and also necessary for the quasi-equilibrium assumption in \S\ref{sec:warm-background}. As seen in Eq.~\eqref{eq_bulk}, the bulk viscosity coefficient $\zeta_b$ corresponds to solely the tracial part of spatial velocity gradient. Thus, bulk stress is related to isotropic pressure fluctuation and already contributes at the background level, contrary to the shear stress which enters only at the perturbation level. Bulk viscosity is typically much smaller than shear viscosity~\cite{Jeon1995,JeonYaffe1996} and is not relevant for the non-trivial evolution of vector metric perturbations which requires the presence of transverse momentum fluctuations $v_i^{V}$.

For a radiation bath that is hot enough to be treated relativistically, effective hydrodynamical transport coefficients such as shear viscosity at interest can be computed using a kinetic theory of relativistic bath particles with physical mass scales $m_{\rm phys}\lesssim\mathcal T$~\cite{BasteroGil:2014,Jeon1995}. The typical functional dependence of the shear viscosity is cubic in temperature,
\begin{equation}
\zeta_s(\eta)=\alpha\mathcal T_0^3,
\label{eq:warm-viscosity}
\end{equation}
with a dimensionless microphysical coefficient $\alpha$. For a weakly coupled scalar theory, for example, $\alpha \simeq 3\times10^3\,\lambda^{-2}$ depends on the microscopic coupling strength $\lambda$~\cite{Jeon1995,JeonYaffe1996}. This model-specific estimate is distinct from the illustrative benchmark adopted below.

\subsection{Introducing thermal stochastic noise}
To maintain the local quasi-equilibrium of the radiation bath and the viscous fluid approximation, damping effects associated with non-vanishing viscous stress need to be balanced by an “outside” stochastic stress in the effective fluid equations, e.g., classical stochastic noises of the radiation bath. In the high temperature regime, we can follow \cite{BasteroGil:2014} and \cite{JeonYaffe1996}, replace deterministic fluid equations by a set of dynamical equations that include thermal noises as sources using Landau’s theory of fluctuating hydrodynamics~\cite{Landau2004,Landau1957} to obtain the power spectrum of small fluctuations around equilibrium such as the vector mode. Specifically, the two-point correlator of the stochastic shear stress can be fixed by the fluctuation-dissipation relation, so that stochastic averages of fluid variables reproduce corresponding thermal ensemble averages with the presence of shear viscous stress. This will be discussed in detail in the next section when we try to compute the vector power spectrum.

We now add this stochastic stress $S^{ij}$ to our previous stress-energy tensor 
\begin{equation}
    T_{ij}=T^{\rm pf}_{ij} + \Pi_{ij}+ S_{ij}.
\end{equation}
Analogously, only the vector part of the noise $S_{ij}^{(V)}$ will contribute to the modified vector mode equation of motion. We extract its transverse amplitude $S(\eta)$ by projecting onto the same Laplacian eigenfunction basis:
\begin{equation}
S_{ij}^{(V)}:=\partial_iS_j+\partial_jS_i\,,\qquad
\partial^iS_i=0\,,\qquad S_i(\mathbf{x},\eta)=S(\eta) Q_i(\mathbf{x})\,,
\label{eq:scalar_noise}
\end{equation}
for each harmonic polarization. The background equations remain invariant upon the introduction of shear noises. Using the full FLRW metric to raise the noise index, the off-diagonal part of the spatial Einstein equation now yields 
\begin{equation}
\omega'+2\mathcal H\omega+\chi+k^2S=0\,.
\label{eq:projected-momentum}
\end{equation}
One can compare this with Eq.~\eqref{eq:without_noise_vector} to see the appearance of the noise term.

Using Eq.~\eqref{eq:reduced-variables} and Eq.~\eqref{eq:momentum-constraint}, with the inertial mass density for radiation $w_r$ kept explicit, the inhomogeneous equation can be put into a compact form 
\begin{equation}
\omega'+A_k(\eta)\omega=-k^2S\,,
\label{eq:omega-eom}
\end{equation}
where
\begin{equation}
A_k(\eta)=2\mathcal H+2\kappa\zeta_s a
+\frac{\zeta_sk^2}{a w_r}\,.
\label{eq:Ak-general}
\end{equation}
There are three distinct deterministic contributions:
\begin{equation}
    A_H=2\mathcal H,\qquad
    A_{F'}=2\kappa\zeta_s a,\qquad
    A_{\rm diff}=\frac{\zeta_s k^2}{a w_r}\,.
    \label{eq:deterministic-damping-terms}
\end{equation}
The shear viscous stress terms $A_{F'}$ and $A_{\rm diff}$ and the Hubble friction term $A_H$ in Eq.~\eqref{eq:deterministic-damping-terms} damp the vector perturbation, or equivalently the transverse momentum flow $\omega$. The term $A_{F'}$ represents local viscous damping due to geometric shear in the metric for modes with $k\neq0$. The wavenumber-dependent term $A_{\rm diff}$ represents damping associated with momentum diffusion~\cite{JeonYaffe1996}. In the local inertial frame, the momentum diffusion coefficient is defined as
\begin{equation}
    D_\zeta \equiv \frac{\zeta_s}{w_r},
    \label{eq:physical-diffusion-definitions}
\end{equation}
and we then have
\begin{equation}
    A_{\rm diff}=aD_\zeta k_{\rm phys}^2\,,
    \label{eq:conformal-diffusion-rate}
\end{equation}
where $k_{\rm phys} \equiv \frac{k}{a}$. On superhorizon scales $A_{\rm diff}$ vanishes as $k/\mathcal H\to0$. The metric shear damping term $A_{F'}$ is still in effect. The kinetic treatment is also automatically justified on superhorizon scales as slow spatial variation requires so that $k_{\rm phys}\ell_{\rm mfp}\ll1$, where $\ell_{\rm mfp}$ is the mean free path of microscopic fluid particles. This suggests, unlike scalar and tensor perturbations, homogeneous vector modes do not freeze at horizon crossing but continue to decay. In Sec.~\ref{sec:sourced-spectrum}, we show that the noise-sourced vector power spectrum in fact also decays in the regime considered here.

We are now in a position to solve this equation with input from background evolution. For constant $\mathcal T_0$, the linear coefficient can be written as a function of $\eta$
\begin{equation}
A_k(\eta)=-\frac{2\beta}{\eta}+c_1(-\eta)^{-\beta}+c_{2}(-\eta)^\beta\,,
\label{eq:Ak-constant-epsilon}
\end{equation}
where
\begin{equation}
c_1=2\kappa\alpha\mathcal T_0^3A_0\,,\qquad
c_{2}=\frac{\alpha\mathcal T_0^3k^2}{A_0w_r}\,.
\label{eq:c1-c2}
\end{equation}
$c_2$ can also be rewritten as
\begin{equation}
    c_2=\frac{45\alpha k^2}{2\pi^2g_*A_0\mathcal T_0},
\end{equation}
using the definition of the radiation inertial mass density $w_r=\frac{2\pi^2}{45}g_*\mathcal T_0^4$. 
The formal solution for $\omega$ can be obtained using the  retarded Green function of Eq.~\eqref{eq:omega-eom}, given by
\begin{equation}
G_k(\eta,\eta')=\Theta(\eta-\eta')
\exp\left[-\int_{\eta'}^\eta A_k(\tau)d\tau\right].
\label{eq:retarded-green}
\end{equation}
Defining
\begin{equation}
d_1=\frac{c_1}{\beta-1},
\qquad
{d_{2}=\frac{c_{2}}{\beta+1},}
\label{eq:d1-d2}
\end{equation}
the Green function becomes
\begin{equation}
\begin{split}
G_k(\eta,\eta')={}&\Theta(\eta-\eta')
\left(\frac{-\eta}{-\eta'}\right)^{2\beta}\\
&\times\exp\left[-d_1\left(( -\eta)^{1-\beta}-( -\eta')^{1-\beta}\right)
+{d_2\left(( -\eta)^{\beta+1}-( -\eta')^{\beta+1}\right)}\right],
\end{split}
\label{eq:constant-epsilon-green}
\end{equation}
given the constant-$\epsilon$ slow roll background dynamics. The sourced particular solution is therefore
\begin{equation}
\omega_{\rm src}(\mathbf{k},\eta)
=-k^2\int_{\eta_0}^{\eta}d\eta'\,
G_k(\eta,\eta') \ S(\mathbf{k},\eta') .
\label{eq:sourced-omega}
\end{equation}
The complementary homogeneous solution evolves the initial momentum flow
\begin{equation}
\begin{split}
\omega_{\rm hom}(\mathbf{k},\eta)
={}&G_k(\eta,\eta_0)\,\omega(\mathbf{k},\eta_0)\\
={}&\left[\frac{a(\eta_0)}{a(\eta)}\right]^2
\exp\left[
-\int_{\eta_0}^{\eta}d\eta'\,
\left(
2\kappa a\zeta_s
+\frac{\zeta_sk^2}{a w_r}
\right)
\right]
\omega(\mathbf{k},\eta_0).
\end{split}
\end{equation}
With no stochastic stress and negligible shear  $A_k(\eta)=A_H$ and clearly the exponential factor reduces to unity, the solution reproduces the homogeneous behaviour for $F'$ using the momentum constraint in Eq.~\eqref{eq:momentum-constraint}\footnote{This non-viscous decay already contains transverse velocity perturbation to begin with. In pure de Sitter vacuum, the momentum constraint sets identically $F'=0$.}
\begin{equation}
\omega_{\rm hom}\propto a^{-2},\qquad F'_{\rm hom}\propto a^{-2},\qquad
\mathcal P_{F',\rm hom}\propto a^{-4}.
\label{eq:homogeneous-decay}
\end{equation}


\section{Noise-Sourced Vector Power Spectrum}
\label{sec:sourced-spectrum}
The primary observable for our analysis is the power spectrum of the gauge-invariant vector metric shear $F_i'$ in synchronous gauge. One can define the dimensionless power spectrum $ \mathcal{P}_{F'}^{(\lambda)} $ per polarization channel of the conformal time derivative $ F'_i $ via the two-point correlation function:
\begin{equation}
\left\langle F'^{(\lambda)}(\mathbf{k},\eta)F'^{(\lambda')}(\mathbf{k}',\eta)\right\rangle=(2\pi)^3\delta^{(3)}(\mathbf{k}+\mathbf{k}')\delta^{\lambda\lambda'}\frac{2\pi^2}{k^3}\mathcal P_{F'}^{(\lambda)}(k,\eta).
\label{eq:spectrum-definition}
\end{equation}
The total spectrum is
\begin{equation}
    \mathcal P_{F'}=\sum_{\lambda=1}^{2}\mathcal P_{F'}^{(\lambda)}\,.
    \label{eq:total-vector-spectrum}
\end{equation}
As mentioned, for the isotropic thermal state considered here, the two polarization channels have equal spectra. It will be implied that spectra and kernels without an explicit polarization label denote a sum over the two transverse directions.

In the previous section, we have established that the momentum constraint in Eq.~\eqref{eq:momentum-constraint} relates the metric vector perturbation to the transverse momentum density flow $\omega$. It is worth reiterating this point: The vector metric mode is not an independent propagating degree of freedom but is tied to the fluid momentum. Consequently, the power spectrum of $F'$ is directly proportional to the power spectrum of $\omega$, defined as
\begin{equation}
\mathcal{P}_{F'}(k,\eta)=\frac{4\kappa^2}{k^4}\mathcal{P}_{\omega}(k,\eta),
\end{equation}
where $\mathcal{P}_{\omega}$ is the dimensionless power spectrum of the rescaled transverse-momentum amplitude, which can be directly computed from the stochastic equation of motion in Eq.~\eqref{eq:omega-eom}.

\subsection{Derivation of the noise kernel}

The stochastic noise $S$ has its origin in the thermal fluctuations of the coarse-grained radiation bath. Following the formalism of fluctuating hydrodynamics~\cite{Landau1957}, we assume different coarse-grained fluid volumes are not correlated with one another and thus introduce a stochastic stress $S_{ab}$ in the local fluid rest frame as an effectively white noise, with $\langle S_{ab}\rangle = 0$ where the brackets denote ensemble average taken over different noise realizations. In the local inertial frame, the fluctuation-dissipation relation fixes the amplitude of the two-point correlator of the stochastic anisotropic stress tensor $S_{ab}$, which we have assumed to be
local in both space and time in the long-wavelength, low-frequency hydrodynamic limit~\cite{Landau2004,BasteroGil:2014},
\begin{equation}
    k_{\rm phys}\ell_{\rm mfp}\ll1,\qquad \omega_{\rm phys}\tau_{\rm therm}\ll1\,.
    \label{eq:noise-hydrodynamic-limit}
\end{equation}
In this limit, the white-noise correlator provides an effective approximation to a potentially coloured correlator with finite microscopic spatial and temporal widths beyond leading-order fluctuating hydrodynamics~\cite{Murase2019}. Latin indices $\{a,b,\ldots\}$ denote orthonormal spatial components in the local inertial frame, which are connected to the comoving coordinate basis $dx^i$  via the vielbein $e^a{}_i$ as $\theta^a=e^a{}_i dx^i$.
\begin{eqnarray}
\left\langle S_{ab}(\mathbf r,t)S_{cd}(\mathbf r',t')\right\rangle
&=&2\mathcal T_0\Big[\zeta_s\left(\delta_{ac}\delta_{bd}+\delta_{ad}\delta_{bc}-\frac23\delta_{ab}\delta_{cd}\right)+\zeta_b\delta_{ab}\delta_{cd}\Big]\nonumber\\
&&\times\delta_{\rm phys}^{(3)}(\mathbf r-\mathbf r')\delta(t-t').
\label{eq:local-noise}
\label{eq:local-noise-full}
\end{eqnarray}
The shear term proportional to $\zeta_s$ is traceless on either index pair while
$\zeta_b$ carries the trace and describes the fluctuating isotropic pressure\footnote{We can use Appendix~\ref{app:lif-flrw} to convert this expression to the stochastic stress $S_{ij}=e^a{}_i e^b{}_jS_{ab}$ in the comoving coordinate basis using the vielbein $e^a{}_i=a(\eta)\delta^a{}_i$, with no net scale factor appearing in this change of basis.}.
The transformation to comoving coordinates leaves the thermodynamic coefficient
completely unchanged, as shown in Appendix~\ref{app:lif-flrw}. Explicitly, the correlator in the comoving coordinate basis is
\begin{eqnarray}
\left\langle S_{ij}(\mathbf x,\eta')S_{mn}(\mathbf x',\eta'')\right\rangle
&=&2\mathcal T_0\Big[\zeta_s\left(\delta_{im}\delta_{jn}+\delta_{in}\delta_{jm}-\frac23\delta_{ij}\delta_{mn}\right)+\zeta_b\delta_{ij}\delta_{mn}\Big]\nonumber\\
&&\times\delta^{(3)}(\mathbf x-\mathbf x')\delta(\eta'-\eta'').
\label{eq:comoving-stress-noise}
\end{eqnarray}
The factor $a^4$ from transforming the two covariant stress tensors cancels the factor $a^{-4}$ from the spatial and temporal delta functions, leaving the same form and in particular the same physical temperature and viscosity coefficients.

After projecting onto the transverse vector sector (see Appendix \ref{app:Vector Noise}), the scalar noise amplitude $S(k,\eta)$ satisfies
\begin{equation}
\left\langle S^{(\lambda)}(\mathbf k,\eta')S^{(\lambda')}(\mathbf k',\eta'')\right\rangle=(2\pi)^3\delta^{(3)}(\mathbf k+\mathbf k')\delta(\eta'-\eta'')\delta^{\lambda\lambda'}\mathcal N_S^{(\lambda)}(k).
\label{noise1}
\end{equation}
Here $S^{(\lambda)}$ denotes the scalar noise amplitude in a fixed polarization channel, as defined in Eq.~\eqref{eq:scalar_noise}.
\(\mathcal N_S^{(\lambda)}\) is defined as the noise kernel per polarization for the projected tensorial noise correlator as in Eq.~\eqref{eq:reduced-scalar-noise-kernel}.
The trace, or bulk-viscosity, part drops out after vector projection.
Physically, this means isotropic pressure fluctuations (bulk noise) cannot source divergenceless vector modes. Only the shear viscosity component contributes.
Hence, for one polarization,
\begin{equation}
\mathcal N_S^{(\lambda)}(k)=\frac{2\mathcal T_0\zeta_s}{k^2}=\frac{2\alpha\mathcal T_0^4}{k^2}.
\label{eq:per-polarization-kernel}
\end{equation}
Thus the total noise kernel is
\begin{equation}
\mathcal N_S(k)=\frac{4\alpha\mathcal T_0^4}{k^2}.
\label{eq:noise-kernel}
\end{equation}
Details on the vector projector $\pi^{lm}_i$ and the derivation for the noise kernel are given in Appendix \ref{app:Vector Noise}. The key point is to note that $ \mathcal{N}_S(k) $ is time-independent in the near-constant temperature approximation, and scales as $k^{-2}$.

\subsection{Vector power spectrum}

We can now  compute the power spectrum using the solution for $\omega$ derived in Sec.~\ref{sec:vector-dynamics}. Neglecting the rapidly-decaying homogeneous solution, the sourced vorticity is given by Eq.~\eqref{eq:sourced-omega}. The two-point correlation function, therefore, takes the form
\begin{equation}
\left\langle\omega^{(\lambda)}(\mathbf k,\eta)\omega^{(\lambda')}(\mathbf k',\eta)\right\rangle
={}k^2k'^2\int_{\eta_0}^{\eta}d\eta'\int_{\eta_0}^{\eta}d\eta''\,G_k(\eta,\eta')G_{k'}(\eta,\eta'')
\times\left\langle S^{(\lambda)}(\mathbf k,\eta')S^{(\lambda')}(\mathbf k',\eta'')\right\rangle .
\label{eq:omega-correlator-double}
\end{equation}
Substituting for the noise correlator from Eq.~\eqref{noise1}, the delta function $\delta(\eta' - \eta'')$ collapses the double integral into a single integral
\begin{equation}
\left\langle\omega^{(\lambda)}(\mathbf k,\eta)\omega^{(\lambda')}(\mathbf k',\eta)\right\rangle=(2\pi)^3\delta^{(3)}(\mathbf k+\mathbf k')\delta^{\lambda\lambda'} k^4\mathcal N_S^{(\lambda)}(k)\int_{\eta_0}^{\eta}d\eta'\,G_k(\eta,\eta')^2 .
\label{eq:omega-correlator-single}
\end{equation}

Using the relation $\mathcal{P}_{F'} = (4\kappa^2/k^4) \mathcal{P}_{\omega} $ and the definition of the dimensionless spectrum $\mathcal{P}_{\omega} = (k^3/2\pi^2) P_{\omega}$, we arrive at the master equation for the vector power spectrum:
\begin{equation}
\mathcal P_{F'}(k,\eta)=
\frac{2\kappa^2k^3}{\pi^2}
\int_{\eta_0}^{\eta}d\eta'\,
G_k(\eta,\eta')^2\mathcal N_S(k,\eta').
\label{eq:master-spectrum-general}
\end{equation}
This is the exact integral form of the sourced power spectrum.  The
white-noise delta function has reduced the two time integrals to one. To explicitly see the extraction of vector noise correlator and the derivation of the dimensionless power spectrum as a time integral, please refer to Appendix \ref{app:Vector Noise}. Substituting the noise kernel in Eq.~\eqref{eq:noise-kernel} into the expression above gives
\begin{equation}
\mathcal{P}_{F'}(k,\eta)=
\frac{8\kappa^2\alpha\mathcal T_0^4k}{\pi^2}
\int_{\eta_0}^{\eta}d\eta'\,[G_k(\eta,\eta')]^2 .
\label{eq:master-spectrum}
\end{equation}
To evaluate the above integral explicitly, we shall substitute the Green's function computed in Eq.~\eqref{eq:retarded-green}. It is convenient to change variables to positive conformal (time) coordinates \begin{equation} x \equiv -\eta',\qquad y \equiv -\eta,\qquad y_0 = -\eta_0. \end{equation} The exact formal
constant-\(\epsilon\) result is
\begin{equation}
\mathcal P_{F'}(k,\eta;\eta_0)=\frac{8\kappa^2\alpha\mathcal T_0^4k}{\pi^2}y^{4\beta}e^{-2d_1y^{1-\beta}+2d_2y^{\beta+1}}\int_y^{y_0}dx\,x^{-4\beta}e^{2d_1x^{1-\beta}-2d_2x^{\beta+1}}\,.
\label{eq:exact-spectrum}
\end{equation}
The exponent $ \beta+1 $ arises from integrating the gradient-viscosity term $ \propto (-\eta)^\beta $ in the damping coefficient $A_k(\eta)$. Note that this integral captures the full history of noise accumulation from the initial time $\eta_0$ to the observation time $\eta$.

The structure of the power spectrum reveals the competition between sourcing and damping. The integral represents a sum of independent noise impulses $S$ injected at each time $\eta'$. Each impulse is weighted by the square of the retarded response $G_k^2$. The factor 
\begin{equation} \left(\frac{y}{x}\right)^{4\beta} = \left[\frac{a(\eta')}{a(\eta)}\right]^4, 
\end{equation} 
represents the standard redshift decay of vector modes due to Hubble damping $A_H$. The exponential terms involving $d_1$ and $d_2$ represent additional suppression due to shear viscosity, corresponding to $A_{F'}$ and $A_{\rm diff}$ respectively. This is a consequence of the additional viscous damping in the model. Finally, the prefactor $k$ in the above result indicates an overall blue tilt when diffusive damping corrections are negligible and the Green's function is independent of $k$. This is consistent with the introduction of viscous stochastic noise, which in general tends to add more power on smaller length scales~\cite{BasteroGil:2014}. In other words, the $k^{-2}$ in the noise kernel partially offsets the phase space factor $k^3$.\begin{equation} \mathcal{P}_{F'} \propto k^3 \mathcal{N}_{S} \propto k. \end{equation}


\section{Late-Time Asymptotic Regimes and Their Validity}
\label{sec:asymptotic-regimes}

The exact power spectrum in Eq.~\eqref{eq:exact-spectrum} involves an integral over the history of different inflationary epochs. This integral admits a pre-asymptotic regime and a formal late-time asymptotic regime depending on the relative magnitude of the Hubble damping term versus the viscous damping terms. Studying its late-time behaviour, we find that shear viscous damping evolves to dominate over Hubble damping and a transition between the pre-asymptotic and the strict asymptotic solutions occurs at a characteristic conformal time $\eta_{\text{turn}} $, or equivalently, a turnover scale $z_{\text{turn}} = |k \eta_{\text{turn}}|$. Let us define $r_\zeta$ as the ratio of geometric-shear damping to Hubble damping and $d$ as the diffusive damping rate in units of the Hubble expansion rate, \begin{equation} r_\zeta \equiv \frac{A_{F'}}{A_H} = \frac{\kappa\zeta_s}{H}, \qquad d \equiv \frac{A_{\rm diff}}{\mathcal H} = \frac{D_\zeta k_{\rm phys}^2}{H}\,. \label{eq:damping-ratios} \end{equation} For a given mode $k=H_*$, $N=\ln a$ counts the number of e-folds the mode spends after horizon crossing with the same normalization condition $a_*=1$ as before. The relative magnitudes evolve as \begin{equation} r_\zeta(N)=r_{\zeta,*}e^{\epsilon N}, \qquad d(N)=d_*e^{-(2-\epsilon)N}\,. \end{equation} The diffusive damping rate $d$ evolves rapidly to vanish on superhorizon scales. The ratio $r_\zeta$ increases slowly as $H$ decreases and geometric-shear damping gradually takes over.  Balancing $A_H$ and $A_{F'}$  yields a practical turnover estimate as
\begin{equation}
z_{\rm turn}=k\left(
\frac{4\kappa\alpha\mathcal T_0^3A_0}{4\beta-1}
\right)^{1/(\beta-1)}\, .
\label{eq:turnover}
\end{equation}
With standard choices of parameters, $N_{\rm turn}$, the number of e-folds the mode has spent after horizon crossing when the well-approximated regime shifts from pre-asymptotic to strict asymptotic is around 840 e-folds, which will be discussed in more detail in Sections~\ref{subsec:turnover-scale} and~\ref{subsec:asymptotic-consistency}. This is clearly fantastically late and physically unrealistic for inflation to occur for. For \(z\gg z_{\rm turn}\), $A_H$ dominates over $A_{F'}$ to shape the dynamics along with the stochastic stress $S_{ij}$. The pre-asymptotic solution Eq.~\eqref{eq:exact-spectrum} is the more accurate
approximation for the full power spectrum. For \(0\lesssim z\ll z_{\rm turn}\),
$A_{F'}$ takes over so that the vector mode settles into an equilibrium balancing both shear dissipation and noise, and the formal asymptotic solution Eq.~\eqref{eq:strict-asymptotic} best approximates the full solution. 

The unusual feature of the power spectrum in this shear-damping dominated regime is the spectrum retains its dependence on $\mathcal T_0$ but the explicit dependence on $\zeta_s$ cancels because both stochastic sourcing and shear damping are proportional to $\zeta_s$. This cancellation is in a way reminiscent of the stationary velocity distribution reached by the solution to the Langevin equation for Brownian motion, which is independent of the friction term~\cite{Risken1996}. Although near the turnover one must return to~Eq.~\eqref{eq:exact-spectrum} for numerical accuracy, we will show that the physically relevant and accessible regime is strictly the pre-asymptotic solution, beyond which the mathematical extrapolation of the vector EOM obtained using a first-order relativistic hydrodynamical approximation no longer supports a physical interpretation in an inflationary background. The detailed mathematical derivation of these limits is provided in Appendix \ref{app:Late-Time Asymptotic}. Here, we focus on their physical implications.


\subsection{The pre-asymptotic Hubble-dominated regime}

In the late time limit, $ \eta \to \eta_{turn}$, the  damping effect due to momentum diffusion $A_{\rm diff}$ is absent. For the late observable inflationary period, Hubble damping due to spacetime expansion $A_H=2\beta y^{-1}$ dominates over the local viscous damping rate  \(A_{F'}=c_1y^{-\beta}\). For the slow-roll regime \(1\lesssim\beta\), the strict asymptotic limit
\(y\to0^+\) at fixed nonzero \(\alpha\) and \(\epsilon\) is eventually
controlled by the local shear-viscous term \(A_{F'}\) as mentioned. In this regime however, the exponential viscous suppression in the retarded Green's function is negligible. The power spectrum evolves in time as
\begin{equation}
\begin{aligned}
\mathcal P_{F'}(k,\eta)
&\approx
\frac{8\kappa^2\alpha\mathcal T_0^4k}
{\pi^2(4\beta-1)}(-\eta) \\
&= \frac{8\kappa^2\zeta_s\mathcal T_0 k A_0^{1/\beta}}
{\pi^2(4\beta-1)}\,a^{-(1-\epsilon)} \,,
\end{aligned}
\label{eq:pre-asymptotic}
\end{equation}
where we have recalled that the scale-factor goes as  $a(\eta) \propto (-\eta)^{-\beta}$. Crucially, this expression is linear in the viscosity coefficient $\alpha$ and equivalently $\zeta_s$. This means that in the physically accessible window, the amplitude of vector modes is directly proportional to the shear viscosity of the radiation bath. This regime persists as long as the geometric shear term in the damping coefficient $ A_k(\eta) $ remains subdominant to the Hubble term. Importantly, note that the power spectrum in this regime decays as $a^{-(1-\epsilon)}$, drastically slower than $a^{-4}$. 


\subsection{The viscosity-dominated formal asymptotic regime}

In the far future limit $ \eta_{turn}>\eta \to 0^-$, the viscous damping term eventually dominates the exponent. The system reaches a stationary equilibrium where noise sourcing balances viscous dissipation. Assuming the slow-roll regime \(1\lesssim\beta\), and at fixed non-zero \(\alpha\), the power spectrum takes the form
\begin{equation}
\mathcal P_{F'}(k,\eta)\simeq
\frac{2\kappa\mathcal T_0k}{\pi^2A_0}(-\eta)^\beta
=\frac{2\kappa\mathcal T_0}{\pi^2}\frac{k}{a(\eta)}\, .
\label{eq:strict-asymptotic}
\end{equation}

In this limit, the explicit dependence on $\alpha$ cancels out. This cancellation occurs because the same transport coefficient $\zeta_s \propto \alpha$ controls both the noise amplitude (fluctuation) and the damping rate (dissipation). This is a manifestation of the fluctuation-dissipation theorem in an expanding background. However, as discussed below, this regime is strictly inaccessible during observable inflation. Let us reiterate that the above cancellation is, however, a formal asymptotic statement. It is only reached exactly when thermal equilibrium is maintained throughout the quasi-dS  inflationary phase and extrapolated indefinitely.  For reasonably small effective vector shear response, the representative regime lies instead in the long pre-asymptotic branch.


\subsection{The turnover scale and observable windows}\label{subsec:turnover-scale}
Since \(\beta-1\simeq\epsilon\ll1\), the transition scale is sensitive to both \(\alpha\) and the slow roll parameter. Consequently, the smaller the effective shear response, the longer the pre-asymptotic regime lasts.  Using the ratio $r_\zeta$ defined in Eq.~\eqref{eq:damping-ratios}, we obtain the formal turnover e-fold $N_{\rm turn}$, measured from horizon crossing with $a_*=1$, as
\begin{equation}
N_{\rm turn}=\frac{1}{\epsilon}\ln\!\left(\frac{3+\epsilon}{4r_{\zeta,*}}\right),
\qquad
r_{\zeta,*}=\frac{\kappa\alpha\mathcal T_0^3}{H_*}.
\end{equation}
For a small viscous damping advantage $r_{\zeta,*}$ and small $\epsilon$, the turnover happens exceptionally late. We can use the following parameter choices for illustrative purposes:
\begin{equation}
\begin{aligned}
&H_*=k=10^{12}\,{\rm GeV},\qquad \mathcal T_0=10^{14}\,{\rm GeV},\\
&M_{\rm Pl}=2.435\times10^{18}\,{\rm GeV},\qquad \kappa=M_{\rm Pl}^{-2},\\
&\alpha=10^3,\qquad g_*=228.75,\qquad \epsilon=10^{-2}.
\end{aligned}
\label{eq:benchmark}
\end{equation}
The reference mode $k$ exits the Hubble radius at $N=0$ when the Hubble scale $H=H_*$ and the scale factor is set to be $a_*=1$. Under these choices and assuming the temperature to be evolving sufficiently slowly, we have $N_{\rm turn}\simeq840.3$, or equivalently $\log_{10}|z_{\rm turn}|\simeq-361.30$. Let us evaluate the power spectrum at an initialized scale $z_0=|k\eta_0|$. The exact choice of the initialized scale does not matter since previous history has negligible influence on the late time behaviour of the vector power spectrum, recalling Eq.~\eqref{eq:homogeneous-decay}. Please also refer to Appendix \ref{app:Late-Time Asymptotic} to see how the initial value drops out from the integral. Let us now compute the integral in Eq.~\eqref{eq:exact-spectrum} initialized at scale $|k\eta_0|=10$ which is slightly before the horizon crossing time.
\begin{figure}[H]
    \centering
    \includegraphics[width=0.8\textwidth]{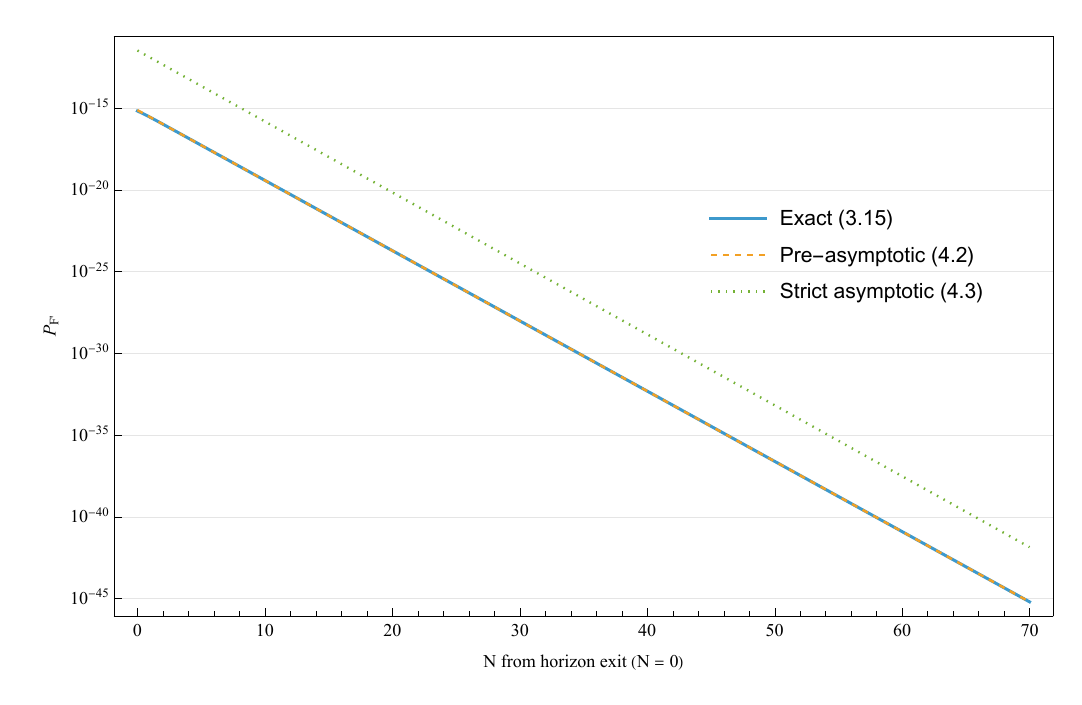}
    \caption{The vector power spectrum is plotted as a function of the number of e-folds $N$ the comoving mode spent after horizon exit ($N=0$), for the benchmark parameter choices listed in Eq.~\eqref{eq:benchmark}. The full exact spectrum is compared with the pre-asymptotic approximation in Eq.~\eqref{eq:pre-asymptotic} and the strict-asymptotic approximation in Eq.~\eqref{eq:strict-asymptotic} over $0\leq N\leq70$.}
    \label{fig:vector-spectrum-horizon-N}
\end{figure}

We find that the pre-asymptotic expression approximates the exact power spectrum Eq.~\eqref{eq:exact-spectrum} impressively accurately, in contrast to the formal asymptotic expression during the physically relevant range $N\lesssim 70$ e-folds.  In fact, within 5 e-folds, the accuracy measured by $\mathcal P_{\rm exact}/\mathcal P_{\rm pre}$ reaches $0.999$. The shear response then enters the physically observable late time limit, the Hubble-dominated regime. Approaching $N_{\rm turn}$, both regimes cease to be good approximations as $\mathcal P_{\rm exact}/\mathcal P_{\rm strict}$ and $\mathcal P_{\rm exact}/\mathcal P_{\rm pre}$ give approximately $0.500416$. Under the assumption of constant temperature, the accuracy markers are listed below in Table~\ref{tab:accuracy-markers}. It is worth noting however the strict asymptotic expression for the vector power spectrum has been extrapolated beyond the validity domain for the assumed inflationary background and first-order hydrodynamical approximations, and is shown here in comparison with the physically accessible pre-asymptotic regime.
\begin{table}[!htbp]\centering
\begin{tabular}{lrr}
\toprule
Ratio & Level & $N$\\
\midrule
$\mathcal P_{\rm exact}/\mathcal P_{\rm pre}$ & $0.99$ & $1.748$\\
$\mathcal P_{\rm exact}/\mathcal P_{\rm pre}$ & $0.99$ & $381.1$\\
$\mathcal P_{\rm exact}/\mathcal P_{\rm pre}$ & $0.50$ & $840.3$\\
$\mathcal P_{\rm exact}/\mathcal P_{\rm strict}$ & $0.90$ & $1059.8$\\
$\mathcal P_{\rm exact}/\mathcal P_{\rm strict}$ & $0.99$ & $1299.5$\\
\bottomrule
\end{tabular}
\caption{Accuracy markers for the pre-asymptotic and strict asymptotic regimes for the benchmark parameters listed in Eq.~\eqref{eq:benchmark}}\label{tab:accuracy-markers}
\end{table}
\begin{figure}[H]
    \centering
    \includegraphics[width=0.7\textwidth]{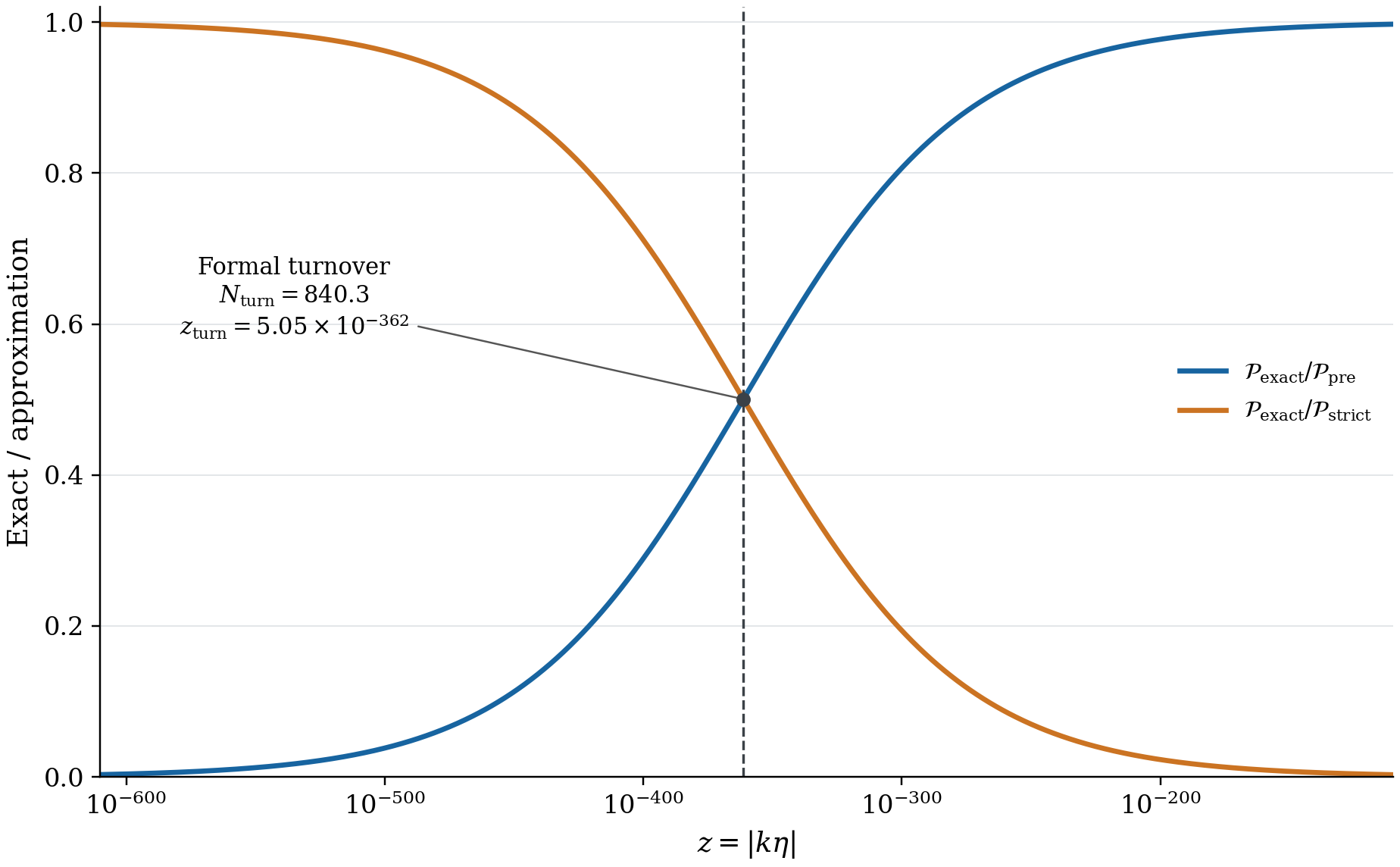}
    \caption{Ratios of the exact full vector power spectrum to the pre-asymptotic and strict-asymptotic expressions are plotted here as functions of $z=|k\eta|$ for the parameter choices given in Eq.~\eqref{eq:benchmark}. The marked formal turnover occurs around $N_{\rm turn}\simeq840.3$, or equivalently $z_{\rm turn}\simeq5.05\times10^{-362}$. }
    \label{fig:vector-accuracy-unscaled-z}
\end{figure}

The Figure above showcases how accurately well-approximated the full power spectrum is by the pre-asymptotic and the formal asymptotic expressions, where the characteristic turnover scale outside the physically observable window visualises the change in terms of the damping effect in domination. We keep only the sourced particular solution, neglecting any vanishingly small initial
homogeneous contributions. For more details on the derivation of the two (pre-)asymptotic regimes, please see the Appendix \ref{app:Late-Time Asymptotic}.

The results in this section again assume a constant-\(\epsilon\) slow-roll background and an approximately
constant physical bath temperature,
\(\mathcal T_0\). The strict
\(\alpha\)-independent result in the formal asymptotic limit is therefore valid only while the physical
temperature remains meaningful as an equilibrium quantity.  After warm
inflation ends, tracking the evolution of the sourced vector modes requires the
subsequent cosmic thermal history. In both late-time regimes considered
here, the gradient-viscosity contribution to the retarded response is
subdominant at leading order, and the resulting spectrum has the simple linear
\(k\)-dependence. For modes deep within the horizon where strong momentum diffusion actively contributes to the vector dynamics or an appreciably evolving physical temperature requires returning to the master expression in Eq.~\eqref{eq:master-spectrum} with input from corresponding background dynamics and updated noise kernel.


\subsection{Causality and consistency bounds}\label{subsec:asymptotic-consistency}

There is a natural way to see that the strict-asymptotic limit, dominated by viscosity, lies out of range of the validity of our approximations. The relative strength between two damping terms of different origins is measured by the previously introduced ratio $r_\zeta=\frac{c_1}{2\beta}y^{1-\beta}
\propto a^\epsilon$ and recall the definition of the momentum diffusion constant in the local inertial frame as $D_\zeta \equiv \zeta_s/w_r$. 
The Maxwell--Cattaneo equation with a finite relaxation time $\tau_\pi$ is a phenomenological step towards the fully covariant Israel--Stewart framework for relativistic causal hydrodynamics~\cite{Romatschke2010,Brito2020IsraelStewart}. The Maxwell-Cattaneo law introduced below is an extension of Navier-Stokes equation that preserves causality and regulates the ill-posed initial value problem in hydrodynamics. Here, the viscous relaxation time $\tau_\pi$ can be considered as a new phenomenological transport coefficient in the local inertial frame~\cite{Romatschke2010,Brito2020IsraelStewart}
\begin{equation}
\tau_\pi\partial_t\Pi_{ij}+\Pi_{ij}
=-2\zeta_s\,\partial_{(i}v_{j)}^V,
\label{MaxwellCattaneo}
\end{equation}
where the local transport coefficients only depend on temperature $\mathcal{T}$ and thus for our purpose can be seen as slowly-evolving constants. We do not consider indeterministic noises since that does not qualitatively change the effective phenomenological analysis here. The Maxwell-Cattaneo theory reproduces the standard acausal diffusive behaviour in the low frequency and low wavenumber limit, i.e. the hydrodynamical limit. This however modifies the dispersion relation and gives rise to effective dynamics of propagating waves in the large wavenumber limit. Together with the momentum conservation equation, subluminal propagation explicitly imposes an upper bound on the characteristic speed of transverse shear perturbations $v_T$
\begin{equation}
    v_T^2=\frac{D_\zeta}{\tau_\pi}
=\frac{\zeta_s}{w_r\tau_\pi}\leq1.
\end{equation}
This is the transverse causality condition for the relaxation model. Using both the background consistency relation $w_r\leq w = 2\epsilon H^2/ \kappa$ and the causality condition $D_\zeta \leq \tau_{\pi}$,
\begin{eqnarray}
    r_\zeta = \frac{\kappa w_r}{H^2} \left( H D_\zeta\right) \leq 2 \epsilon \left( H D_\zeta\right)\,<2 \epsilon \left( H \tau_{\pi}\right).
\end{eqnarray}
Writing $\tau_\pi=C_\pi D_\zeta$, where $C_\pi\equiv v_T^{-2}$ depends on the bath microphysical couplings and can be treated as a constant, the first-order hydrodynamic approximation on an expanding spacetime background requires the following condition
\begin{equation}
H\tau_\pi
=C_\pi\frac{45\alpha}{2\pi^2g_*}\frac{H}{\mathcal T_0}\ll1,
\label{eq:hydrodynamic-relaxation-condition}
\end{equation}
i.e., the relaxation time must be short compared to the expansion timescale so that stress relaxation can be eliminated at the level of the slow response in Eq.~\eqref{MaxwellCattaneo}. This is qualitatively equivalent with imposing the condition $H\ell_{\rm mfp}\ll1$ to ensure the validity of the hydrodynamical EFT on an expanding background~\cite{QiuSorbo2021}. In this regime, the deterministic shear stress approaches its Navier--Stokes value, while $S_{ij}$ reduces to the white stress used in the first-order calculation. For the chosen benchmark parameters, $H_*D_\zeta\simeq0.09966$ and $H_*\tau_\pi\simeq0.09966\,C_\pi$. The exact value of constant $C_\pi$ does not qualitatively change our discussion. Following \cite{Brito2020IsraelStewart} and taking the typical value $C_\pi=5$ in the ultrarelativistic limit~\cite{YorkMoore2009} gives $H_*\tau_\pi\simeq0.4983$ within the stated limit. A smaller prefactor $\alpha$ or constant $C_{\pi}$ will give a clearer separation between the relaxation and expansion timescales, which does not qualitatively change the physical conclusions made here. It is also easy to see that the condition $H\tau_{\pi}$ decreases with time since $H(N)=H_*e^{-\epsilon N}$, given that the cosmological background is still well described by slow-roll inflationary expansion with a background radiation bath, and thus the upper bound continues to be satisfied for the subsequent expansion in a physically tenable model. The general relaxation requirement also automatically implies 
\begin{equation}
    r_\zeta \ll \epsilon,
\end{equation}
throughout the regime in which we want to apply the hydrodynamic approximation. This means that the local viscous damping rate cannot overtake Hubble damping without leaving the theory's own domain of validity. For example, at the reference time $\eta_*$, $r_{\zeta,*}\simeq1.68656\times10^{-4}$ and although this value will slowly increase consistent with the slowly happening transition from the Hubble-dominated regime to the geometric shear dominated regime, it remains strictly subdominant to the slow-roll parameter within the physically observable window. Therefore, the asymptotic regime, where Eq.~\eqref{eq:strict-asymptotic} is valid, is  not merely very late, but physically inaccessible under the assumptions of the model.

For the same benchmark, the ratio of the radiation inertial mass density to the total inertial mass density at the reference time is $w_{r,*}/w_*\simeq0.0846156$. Under the constant-$\mathcal T_0$, constant-$\epsilon$ extrapolation of the vector mode solution, $w_r$ remains constant while $w(N)=w_*e^{-2\epsilon N}$, so that the radiation enthalpy reaches the total enthalpy in the universe, $w_r=w$, at $N\simeq123.5$, already well before the formal turnover at $N\simeq840.3$. Thus the background dynamics consistency bound on radiation enthalpy, $w_r<w$ independently limits the period of physical applicability of the hydrodynamical calculation given here to $N\simeq120$. Despite being much earlier than when the first-order hydrodynamical analysis ceases to apply, this suggests the pre-asymptotic expression remains the only physically accessible regime given that observable scales exit the horizon only $\sim50$--$60$ e-folds before the end of inflation~\cite{Baumann2022}. In the superhorizon limit, another hydrodynamical consistency condition to ensure slow spatial variation of the macroscopic hydrodynamical properties, namely $k_{\rm phys}\ell_{\rm mfp}\ll1$, where $\ell_{\rm mfp}$ is the mean free path of microscopic fluid particles, is also comfortably satisfied.  As a further check, the shear viscosity to volume entropy density ratio for the benchmark values is
\begin{equation}
    \zeta_s/s = \frac{45\alpha}{2\pi^2 g_*} \simeq 9.97\,,
\end{equation}
where $s=w_r/\mathcal T_0=2\pi^2g_*\mathcal T_0^3/45$. This clearly exceeds the conjectured universal KSS lower bound $1/(4\pi)$~\cite{KovtunSonStarinets2005, IqbalLiu2009} and is consistent with the large viscosity-to-entropy ratio expected in weakly coupled relativistic quantum field theories at finite temperature and vanishing chemical potential~\cite{Jeon1995,JeonYaffe1996}.

\section{Discussion}
\label{sec:discussion}
\subsection{Validity of superhorizon noises}
A potential concern is whether we can safely apply the first-order fluctuating viscous hydrodynamical approximation on superhorizon scales ($k \ll aH$). In other words, can we still meaningfully treat the radiation bath as a viscous fluid once a perturbation's wavelength has grown much larger than the Hubble horizon?  As shown in the previous section, both sufficiently slow temporal variation $H\tau_\pi\ll1$ and spatial variation $k_{\rm phys}l_{\rm mfp}\ll1$ were satisfied in our calculation to justify the application of first-order dissipative relativistic hydrodynamics. Understanding the fluctuating hydrodynamics theory as an effective field theory, $l_{\rm mfp}$ assumes the role of UV cutoff in the derivative expansion, which introduces scale sensitivity in the opposite end of the superhorizon limit which probes large scale cosmological structures \cite{Kovtun2012}. From Eq.~\eqref{eq:omega-eom} and Eq.~\eqref{eq:Ak-general}, we see explicitly that the geometric-shear induced damping $A_{F'}$ is still present when $k\rightarrow0$ as a manifestation of the momentum constraint in Eq.~\eqref{eq:momentum-constraint} on $F'$ derived from linear Einstein equation. Only the diffusive damping term, which explicitly probes spatial variation on the perturbation's own wavelength, switches off as $k/(aH) \to 0$. 

In our calculation, vector dynamics is influenced by both shear damping effects and stochastic noise sources, even on superhorizon scales. It is often implied that because spatial correlations between causally disconnected regions are unphysical, there cannot be a non-vanishing power spectrum at sufficiently large scales~\cite{HuSpergelWhite1997, DurrerCaprini2003}. However, a white stochastic noise in the standard Landau-Lifshitz formulation~\cite{Landau1957,Landau2004} does not lead to any inconsistency with the principle of causality, as it models precisely the lack of correlation between local observables across different coarse-grained fluid volumes. We can motivate this more carefully as follows. The stochastic anisotropic stresses $S_{ij}(x_i,\eta)$ we invoke to source the vector power spectrum in Eq.~\eqref{eq:local-noise} is a local white noise with an amplitude equal to local equilibrium hydrodynamical quantities, namely the physical temperature $\mathcal{T}_0$ and the shear viscosity $\zeta_s$ itself, directly following the Landau–Lifshitz formulation. This is, in principle, applicable on any scale larger than $l_{\rm mfp}$ where deviations from the hydrodynamical EFT approximation kick in. According to Paley–Wiener–Schwartz theorem~\cite{Hormander2003}, any causal correlation, with compact support in real space (white noises only have support on a point), has an analytic Fourier transform in momentum space~\cite{Durrer:1997bs, DurrerCaprini2003}. In small $k$ limit, analyticity implies the power spectrum approaches a constant \[P_{\rm source}(k)=P_0+O(k^2),\] since all higher order terms in $k$ vanish on superhorizon scales. When the constant $P_0$ does not vanish, causal physical processes leave behind a low-$k$ white tail. Indeed, in our case the amplitude of the Fourier transformed two-point noise correlator is explicitly fixed by fluctuation-dissipation relation to be $2\mathcal{T}_0\zeta_s$, a constant under the quasi-equilibrium assumption in warm inflation. It is worth noting that this does not require the vector perturbations to be thermalised, and they are indeed  not assumed to obey a Bose-Einstein distribution. Non-analyticity in $k$ can arise, however, as an artifact of non-local projectors involving spatial derivatives onto specific sectors, such as the noise kernel $\mathcal{N}_S\propto k^{-2}$ computed here via transverse projection and scalar extraction from anisotropic noise $S^{(V)}_{ij} = -i(k_iS_j + k_jS_i)$. This non-analyticity of SVT projectors is responsible for a similar divergence in the scalar anisotropic stress potential on large scales as pointed out by~\cite{Durrer:1997bs}. 

Thus,  a central issue is how to correctly impose causality and identify which object behaves as a white noise in the small $k$ limit~\cite{DurrerMelchiorri2026}. In several studies on causal seeds such as topological defects,~\cite{Magueijo1996, HuSpergelWhite1997, DeruelleLangloisUzan1997}, the scalar shear stress potential $\Pi^s$ defined as $\Pi_{ij} \supset (k_ik_j - \tfrac13\delta_{ij}k^2)\Pi^s$, instead of the stress $\Pi_{ij}$ itself, has been considered to behave as a white noise on large scales $\propto k^0$. This assumption leads to the erroneous conclusion that the corresponding scalar anisotropic stress $\Pi^s_{ij}\propto k^2$ outside the horizon, which has been pointed out by~\cite{Durrer:1997bs} to be inconsistent with large-$N$ limit and numerical simulation results. Large scale power spectra do not have to be highly suppressed to comply with the causality principle. In our vector calculation in warm inflation, the indeterministic stochastic noise $S_{ij}$ represents random kicks between coarse-grained fluid elements in the radiation bath on scales $k_{\rm phys}l_{\rm mfp}\ll1$. It is not a spatial gradient of any long-wavelength field and is itself assumed to be uncoloured in a naturally physically motivated way. 

The large-scale white noise in “kurvature density” \(\Delta\rho\) defined in~\cite{BarenboimStebbins2026} turns out not to result in an infrared-enhanced adiabatic comoving curvature perturbation \(\mathcal R_2\), due to local conservation laws constraining $\mathcal R$ in the soft limit as found by subsequent development~\cite{Ireland2026,DurrerMelchiorri2026}. Analogously, the white stochastic stress $S_{ij}$ of fluctuating hydrodynamics considered here does not leave a conserved superhorizon vector mode either. The momentum constraint ties $F'$ directly to the transverse momentum density flow $\omega$ and dilutes the vector mode via $A_{F'}$ despite the continuous replenishment of noises. The gauge-invariant vector mode $F'$ in our calculation, or equivalently $\omega$, does not have a white noise power spectrum but $|F'_k|^2 \propto k^{-2}$ or equivalently $\langle|\omega_k|^2\rangle \propto k^4\,\mathcal N_S \propto k^2$.

\subsection{More on large scale white noise}
In~\cite{BarenboimIrelandStebbins2025, BarenboimIrelandStebbins2026PRL}, large scale white noise is furthermore argued to be an inevitable consequence of local nonlinear dynamics, where power is transferred to larger scales from unresolved nonlinear small-scale physics. These dynamically generated contributions appear effectively stochastic on large scales. In warm inflation, internal bath dynamics specified by generically nonlinear microphysical couplings is accounted for by introducing stochastic noises after coarse graining.\footnote{Nonlinear interactions among freely propagating hydrodynamic modes also produce corrections to the "bare" values of transport coefficients such as the shear viscosity $\zeta_s$~\cite{Kovtun2012}.} The stochastic stresses $S_{ij}$ from a quickly equilibrating thermal bath in this sense share their fundamental origin with the large scale white noise identified in~\cite{BarenboimIrelandStebbins2025,BarenboimIrelandStebbins2026PRL}. Both can be traced back to unresolved nonlinear short wavelength modes and appear as uncorrelated Poisson-like noises on scales much larger than the characteristic correlation lengths, although in fluctuating hydrodynamics the noise directly appears as a part of the effective description after integrating out bath degrees of freedom. In earlier stochastic inflation discussions~\cite{Matacz1997},  a related open-system perspective is adopted to interpret large-scale inflaton fluctuations as a system subject to the environment made of small-scale fluctuations. Integrating out environmental degrees of freedom yields an effective description incorporating both noise and dissipation. Notably, the bipartition scale can itself be well outside the horizon $k_{min} \ll Ha(t)$. Large scale structure can then be studied as an open quantum system \cite{Shandera}.

A stochastic gravitational-wave background has also been proposed as a source of large-scale white noise through nonlinear short mode coupling~\cite{BarenboimStebbins2026}. In the high-frequency limit of gravitational waves, short-wavelength modes $\sim\lambda_{\rm GW}$ can be averaged over certain spacetime regions at scales $\lambda_{\rm GW}\ll \ell\ll L_{\rm bg}$ to produce an effective stress-energy tensor on the slowly varying background geometry with curvature scale $\sim L_{\rm bg}$ ~\cite{Isaacson1968b}. This coarse-graining process connects naturally to the semiclassical formulation of stochastic gravity~\cite{HuVerdaguer2008}, where stress-energy fluctuations of quantum matter fields in curved spacetime act as the stochastic source in the Einstein--Langevin equation~\cite{HuVerdaguer2008}. The vector perturbation framework developed here opens the possibility to probe the effects of recently emphasized large scale white noise~\cite{BarenboimIrelandStebbins2026PRL,BarenboimIrelandStebbins2025} in a vector channel in warm inflation.

\subsection{Differences between vector and scalar modes}
Another distinctive feature of the vector metric perturbation is that it continues to decay after crossing the horizon, exhibiting a fundamental difference from scalar or tensor modes~\cite{Weinberg2008, Baumann2022, MalikWands2009}. This is already present in the absence of shear and accompanying stochastic anisotropic stresses in the homogeneous vector evolution, Eq.~\eqref{eq:homogeneous-decay}. Similarly sourcing either scalar or tensor metric perturbations by thermal fluctuations does not alter their signature freezing-out behaviour. It has been shown in \cite{HallMossBerera2004,BasteroGil:2014} and \cite{QiuSorbo2021} that scalar and tensor perturbations still approach a constant on superhorizon scales even considering random kicks from thermal noises. To better understand this, let us take the scalar mode as an example and revisit the Mukhanov--Sasaki equation for gauge-invariant comoving curvature perturbation $\mathcal{R}$~\cite{Weinberg2008,Baumann2022},
\begin{equation}
\mathcal R_k''+2\frac{z'}{z}\mathcal R_k'+k^2\mathcal R_k=0,
\label{eq:MS}
\end{equation}
where $z=\frac{a\dot\phi}{H}$. On superhorizon scales, the restoring force \(k^2\mathcal R_k\) becomes negligible, the general solution $\mathcal R_k=C_{1}+C_2\int^\eta\frac{d\eta'}{z^2}$ admits two independent branches, namely the constant adiabatic mode and a decaying mode. Discarding the decaying mode, the comoving scalar curvature mode $\mathcal{R}$ is often said to `freeze out'.

Adiabatic modes are defined to be cosmological perturbations that are locally indistinguishable from a (large) change of coordinates~\cite{Weinberg2003}. An intuitive way to see under what circumstances are adiabatic scalar modes indeed conserved at large scales is the so-called "separate universe" approximation~\cite{WandsMalikLythLiddle2000,LythMalikSasaki2005}. An adiabatic scalar perturbation on spatial hypersurfaces of uniform density can be interpreted to be due to the relative differences in local temporal shift.
Each superhorizon patch can be thought of as evolving independently as a nearby FLRW solution with a slightly
shifted time origin. For an adiabatic scalar perturbation, the curvature
perturbation obeys on large scales~\cite{WandsMalikLythLiddle2000,LythWands2003}
\begin{equation}
\zeta_k'
=
-\frac{\mathcal H}{\rho+p}\,\delta p_{{\rm nad},k}
+\mathcal O\!\left(\frac{k^2}{\mathcal H^2}\right)\mathcal H\Phi_k.
\label{eq:zeta-conservation}
\end{equation}
Once the non-adiabatic pressure becomes negligible, the right-hand side
vanishes as \(k/\mathcal H\to0\), and \(\zeta_k\) approaches a constant.
The relative integrated expansion between the patches is a curvature perturbation directly protected by local energy-momentum conservation. This is why scalars freeze out after crossing the horizon, following the key assumption that on large unobservable scales different patches can be considered to be separately evolving. In~\cite{TanakaUrakawa2021}, the same argument is extended to accommodate non-vanishing large scale shear and applied to construct conserved adiabatic tensor modes. It is clear however the transverse vector mode cannot be argued to freeze in the same way at least, since the gauge-invariant vector mode is $F'$ rather than $F$, which is forced to dilute due to the momentum constraint.  

In contrast to~Eq.~\eqref{eq:MS}, the vector equation of motion in Eq.~\eqref{eq:without_noise_vector} is first order in nature and admits only one branch of homogeneous solution. The constraint in Eq.~\eqref{eq:momentum-constraint} ties the vector metric modes to the momentum density, which continues to redshift
and relax after horizon crossing. In the late-time limit, diffusive damping is highly suppressed so that the superhorizon Green function is effectively independent of $k$, Eq.~\eqref{eq:pre-asymptotic} gives $\mathcal P_{F'}\propto k$. Because $G^2\propto(\eta/\eta')^{4\beta}$ with $4\beta>1$, the time integral is dominated by noise injected within the last few e-folds, so the spectrum carries no memory of horizon crossing. The overall spectral tilt is a combination of the phase-space factor $k^3$ in the dimensionless power spectrum and the vector noise kernel $\propto k^{-2}$ from stochastic anisotropic stress $S^{(V)}_{ij}$. Relative to a causal white-noise spectrum $\mathcal P\propto k^3$, $\mathcal{P}_{F'}$ is therefore IR-enhanced.

It is proposed that exotic models such as solid inflation hosting superhorizon anisotropic stress due to a shear modulus~\cite{EndlichNicolisWang2013, PajerJazayeri2018} could qualitatively change the redshift of the decaying vector mode. In solid inflation, the EOM is second order and contains a restoring force term just as the scalars. The vector mode consequently has the same leading time dependence as scalars. Our result in warm inflation directly shows a significant delay of decay can be achieved as a natural noise-sourced solution in the presence of a background radiation bath, where at superhorizon scales anisotropic stress is present precisely reflecting the dynamical nature of the background bath.

The vector-to-scalar ratio can be defined to be $R_{F'/s}(k,N)
\equiv\frac{\mathcal P_{F'}(k,N)}
{\mathcal P_s(k,N)}$, the ratio between power spectra of the same mode at the same time. From Planck observational bounds~\cite{Planck2018VI}, we see that $P_s\approx2.1\times10^{-9}$ at the pivot scale $k_p=0.05\,\mathrm{Mpc}^{-1}$ where $a_{\rm today}=1$. Setting $a_*=1$ at the horizon crossing time for the pivot scale and converting the different normalizations, we can use Eq.~\eqref{eq:pre-asymptotic} and obtain the numerical value for the vector-to-scalar ratio $N$ e-folds after the pivot mode crossed the horizon
\begin{equation}
R_{F'/s}(k_p,N)\simeq
\frac{8\alpha}{\pi^2(3+\epsilon)P_s}
\left(\frac{\mathcal T_0}{M_{\rm Pl}}\right)^4
e^{-(1-\epsilon)N},
\end{equation}
which is independent of $H_*$.
Evaluated at $N=60$ e-folds after the pivot mode crosses the horizon for example, 
\begin{equation}
R_{F'/s}(k_p,60)\simeq5.82\times10^{-33},
\end{equation}
given that the Hubble energy scale satisfies the previously stated hydrodynamical consistency relations. The ratio is small as expected. Physical observable signatures of the effect proposed in this work will be through second-order effects such as gravitational wave backreaction.

\subsection{Second-order effects} 

The small late-time vector-to-scalar ratio found above indicates that the observable signature for noise-sourced vector modes is likely not a direct detection, but inferred from secondary effects generated when vector modes are still of appreciable magnitude and interacting with other cosmological perturbations beyond the linear order. Nonlinear mode coupling of density fluctuations has already been shown to generate vector metric perturbations during the radiation-dominated era~\cite{LuAnandaClarkson2008}, including higher order effects studied in~\cite{ChangZhangZhou2022}. These results motivate extending the present calculation beyond linear order to study interactions between different sectors. Other possible indirect observable signatures range from vector-induced gravitational waves~\cite{BhaumikPapanikolaouGhoshal2025}, CMB temperature anisotropies and B-mode polarization~\cite{Lewis2004}. These existing observational channels indicate that further exploration is necessary to establish connections to a detectable CMB or gravitational-wave signal.

\section{Conclusions}
This paper has shown that the presence of a radiation bath and associated shear viscous effects during inflation can induce vector mode production. Vector modes obey a constrained Langevin equation sourced by stochastic noises during warm inflation. The formal late-time behaviour of the vector power spectrum is stationary since the same transport coefficient controls noise and damping and the explicit \(\alpha\)-dependence cancels.  In the physically accessible pre-asymptotic regime however, Hubble damping dominates the integral and the spectrum remains linear in \(\alpha\). Remarkably, we find that the standard redshift $\propto a^{-4}$ of vector modes is slowed down significantly by considering both dissipative processes associated with shear viscosity in the radiation bath and accompanying stochastic fluctuations. Post-inflationary vector-mode evolution, however, remains a separate problem which requires further work.

In a dynamical out-of-equilibrium model, viscosity arises within a particle bath from interactions amongst its constituents. A key feature of cold inflation is that inflationary expansion occurs in a supercooled thermodynamic state, so with no production of particles from particle physics interactions (there can be a low level of particle production from cosmological expansion with energy density around or below $\rho_p \sim H^4$, where $\rho_p$ is the energy density of a generic particle bath which may or may not behave as radiation, for a review please see Ford~\cite{Ford2021}). The defining feature of warm inflation is that there is particle production concurrent with inflationary expansion with the particle energy density well above the Hubble scale, $\rho_p \gg H^4$~\cite{BereraFang1995, Berera1995, BereraMossRamos2009}. At this scale, shear viscous effects including both viscous damping and stochastic sourcing can actively participate in vector dynamics~\cite{BasteroGil:2011Shear, BasteroGil:2014}. It is exactly this intrinsically dynamical setting of warm inflation that slows down the decay of vector modes. Viscous shear stress transfers momentum between neighbouring fluid elements and is thus intimately related to momentum diffusion in both flat and curved spacetime. Shear relaxation is a thermodynamically irreversible process and damps vector modes on its own. Embedding the vector mode evolution in the presence of a radiation bath indeed subjects it to shear viscous damping, but also simultaneously immerses vector modes in a bath of classical statistical correlations, \textit{i.e.} stochastic noises. It is exactly this dynamical duo that induces non-trivial vector redshift in joint force. A fundamentally ``warm'' calculation has been presented in this work, where dissipative damping and stochastic noises are paired from the onset, both being fixed by the same shear viscosity specified by the macroscopic environment through the fluctuation--dissipation relation.

We have seen in this work that the direct production of v-modes is suppressed and such modes decay away, leaving no directly observable effect. However this paper establishes that vector modes are present during warm inflation, in sharp contrast to cold
inflation, and such modes potentially can then induce secondary effects. These need to be understood next to arrive at what observable consequences there are from vector modes for warm inflation.  Any second-order effect induced by the non-trivial redshift of vector modes will serve precisely to differentiate between cold and warm inflationary phenomenology. This will be studied in future work.

\appendix
\section{Local inertial frame and comoving-coordinate conventions}
\label{app:lif-flrw}

The harmonic equation in Eq.~\eqref{eq:vector-harmonics} uses the comoving Laplacian
\begin{equation}
\nabla^2\equiv\delta^{ij}\partial_i\partial_j,
\label{eq:conformal-laplacian}
\end{equation}
whose eigenvalue \(k^2\) is time independent.  This operator is different from the Laplace--Beltrami operator $\Delta_\gamma$ constructed from the physical FLRW spatial
metric \(\gamma_{ij}=a^2\delta_{ij}\).  To relate the two, let us introduce the
orthonormal coframe and frame
\begin{equation}
\theta^a=e^a{}_i dx^i=a\delta^a{}_i dx^i,
\qquad
e_a=e_a{}^i\partial_i=a^{-1}\delta_a{}^i\partial_i .
\label{eq:tetrad-basis}
\end{equation}
Then
\begin{equation}
\Delta_\gamma\equiv\gamma^{ij}D_iD_j
=\delta^{ab}e_ae_b
=a^{-2}\delta^{ij}\partial_i\partial_j
=a^{-2}\nabla^2 .
\label{eq:laplacian-translation}
\end{equation}
Here, \(k\) is time independent and is therefore the comoving wavenumber rather than the physical wavenumber is \(k_{\rm phys}=k/a\).  The factor
\(a^{-2}\) in the Einstein-tensor component in Eq.~\eqref{eq:einstein-tensor-momentum} is precisely the conversion displayed above.

The same basis transformation brings the noise correlator in the local inertial frame (LIF) to comoving coordinates.  The frame and coframe conventions are the same as Eq.~\eqref{eq:tetrad-basis}
so that
\begin{equation}
\gamma_{ij}=e^a{}_i e^b{}_j\,\delta_{ab}=a(\eta)^2\delta_{ij}.
\end{equation}
For a spatial covariant rank-$(0,2)$ tensor, one therefore has
\begin{equation}
S_{ij}=e^a{}_i e^b{}_j\,S_{ab}
=
a(\eta)^2\,\delta^a{}_i\delta^b{}_j\,S_{ab}.
\end{equation}
In the LIF, the noise correlator is~\cite{BasteroGil:2014}
\begin{equation}
\begin{split}
\left\langle
S_{ab}(\mathbf r,t')\,S_{cd}(\mathbf r',t'')
\right\rangle
=
\delta_{\rm phys}^{(3)}(\mathbf r-\mathbf r')\,\delta(t'-t'')
\\
\times
2\mathcal T_0
\Big[
\zeta_s\big(\delta_{ac}\delta_{bd}+\delta_{ad}\delta_{bc}\big)
+
\Big(\zeta_b-\frac{2}{3}\zeta_s\Big)\delta_{ab}\delta_{cd}
\Big].
\end{split}
\end{equation}
Given
\begin{equation}
\mathbf r=a(\eta)\,\mathbf x,
\qquad
dt=a(\eta)\,d\eta,
\end{equation}
the Dirac delta functions become
\begin{equation}
\delta_{\rm phys}^{(3)}(\mathbf r-\mathbf r')
=
a(\eta')^{-3}\delta^{(3)}(\mathbf x-\mathbf x'),
\qquad
\delta(t'-t'')
=
a(\eta')^{-1}\delta(\eta'-\eta'').
\end{equation}
The Kronecker deltas transform as
\begin{equation}
\begin{aligned}
e^a{}_l e^b{}_k e^c{}_m e^d{}_n
\big(\delta_{ac}\delta_{bd}+\delta_{ad}\delta_{bc}\big)
&=
\gamma_{lm}\gamma_{kn}+\gamma_{ln}\gamma_{km}
\\
&=
a(\eta')^4
\big(\delta_{lm}\delta_{kn}+\delta_{ln}\delta_{km}\big),
\end{aligned}
\end{equation}
and similarly
\begin{equation}
e^a{}_l e^b{}_k e^c{}_m e^d{}_n\,\delta_{ab}\delta_{cd}
=
\gamma_{lk}\gamma_{mn}
=
a(\eta')^4\delta_{lk}\delta_{mn}.
\end{equation}
Therefore the $a(\eta')^4$ from the tensor conversion cancels the $a(\eta')^{-4}$ coming from the spatial and temporal delta functions, yielding the comoving-coordinate correlator
\begin{equation}
\begin{split}
\left\langle
S_{lk}(\mathbf x,\eta')\,S_{mn}(\mathbf x',\eta'')
\right\rangle
=
\delta^{(3)}(\mathbf x-\mathbf x')\,\delta(\eta'-\eta'')
\\
\times
2\mathcal T_0
\Big[
\zeta_s\big(\delta_{lm}\delta_{kn}+\delta_{ln}\delta_{km}\big)
+
\Big(\zeta_b-\frac{2}{3}\zeta_s\Big)\delta_{lk}\delta_{mn}
\Big].
\end{split}
\end{equation}
Fourier transformation then gives
\begin{equation}
\begin{split}
\left\langle
S_{lk}(\mathbf k,\eta')\,S_{mn}(\mathbf k',\eta'')
\right\rangle
=
(2\pi)^3\delta^{(3)}(\mathbf k+\mathbf k')\,\delta(\eta'-\eta'')
\\
\times
2\mathcal T_0
\Big[
\zeta_s\big(\delta_{lm}\delta_{kn}+\delta_{ln}\delta_{km}\big)
+
\Big(\zeta_b-\frac{2}{3}\zeta_s\Big)\delta_{lk}\delta_{mn}
\Big].
\end{split}
\end{equation}
This is the noise correlator in comoving coordinates with the thermodynamic prefactors kept as physical quantities. In particular, the temperature $\mathcal{T}_0$ is the physical temperature during warm inflation. The physical shear viscosity is 
\begin{equation}
\zeta_s=\alpha\,\mathcal T_0^3.
\end{equation}
Thus we see explicitly how \(\mathcal T_0\) and the viscosities in the main calculation remain
physical scalar quantities, and the projected kernel in Eq.~\eqref{eq:noise-kernel} contains no
additional scale-factor dependence.

\section{The vector noise correlator and power spectrum}
\label{app:Vector Noise}
The projected $\omega$ correlator is linked to the stochastic tensor noise correlator through the Green-function solution for $\omega$:
\begin{equation}
\label{eq:omegaomega-projected}
\begin{aligned}
\left\langle
\omega_i(\mathbf k,\eta)\,\omega^i(\mathbf k',\eta)
\right\rangle
=
&k^4
\int_{\eta_0}^{\eta} d\eta'
\int_{\eta_0}^{\eta} d\eta''\,
G_k(\eta,\eta')\,G_{k'}(\eta,\eta'')
 \\
&\times
\pi^{lk}{}_{i}(-i\mathbf k)\,\pi^{i\,mn}(-i\mathbf k')\,
\left\langle
S_{lk}(\mathbf k,\eta')\,S_{mn}(\mathbf k',\eta'')
\right\rangle .
\end{aligned}
\end{equation}
After projection onto the vector sector and contraction of the remaining tensor indices, define the reduced scalar kernel by
\begin{equation}
\pi^{lk}{}_{i}(-i\mathbf k)\,\pi^{i\,mn}(-i\mathbf k')\,
\left\langle
S_{lk}(\mathbf k,\eta')\,S_{mn}(\mathbf k',\eta'')
\right\rangle
=
(2\pi)^3\delta^{(3)}(\mathbf k+\mathbf k')\,\delta(\eta'-\eta'')\,
\mathcal N_S(k,\eta').\label{eq:reduced-scalar-noise-kernel}
\end{equation}
With this definition, the projected $\omega$ correlator immediately yields the total sourced spectrum formula above.

Following the vector-projector construction of Ref.~\cite{ChangZhangZhou2022}, define the transverse projector in the comoving coordinate basis by
\begin{equation}
\mathsf P^i{}_j(\mathbf k)
:=
\delta^i{}_j-\frac{k^ik_j}{k^2},
\end{equation}
For a symmetric tensor source of vector type
\begin{equation}
S_{lm}^{(V)}(\mathbf k)
=
\!
-i\left(k_lS_m(\mathbf k)+k_mS_l(\mathbf k)\right),
\qquad
k^mS_m(\mathbf k)=0.
\end{equation}
This is the comoving-coordinate version of the vector-sector extraction used in Ref.~\cite{ChangZhangZhou2022}. In the polarization basis of that reference, the same projection is written by contracting the symmetric tensor source with a transverse polarization vector.
The corresponding vector projector can be written as
\begin{equation}
\pi^{lm}{}_{i}(-i\mathbf k)
=
\frac{i}{2k^2}
\left(k^l\mathsf P^m{}_{i}+k^m\mathsf P^l{}_{i}\right).
\end{equation}
It satisfies
\begin{equation}
\pi^{lm}{}_{i}(-i\mathbf k)\,S_{lm}^{(V)}(\mathbf k)=S_i(\mathbf k),
\qquad
\delta_{lm}\,\pi^{lm}{}_{i}=0.\label{eq:vector-source-extraction}
\end{equation}
Explicitly,
\begin{equation}
\begin{aligned}
\pi^{lm}{}_{i}(-i\mathbf k)\,S_{lm}^{(V)}(\mathbf k)
&=
\frac{i}{2k^2}
\left(k^l\mathsf P^m{}_{i}+k^m\mathsf P^l{}_{i}\right)
\Big[-i\left(k_lS_m(\mathbf k)+k_mS_l(\mathbf k)\right)\Big]
\\
&=
\frac{1}{2k^2}
\Big(k^2\mathsf P^m{}_{i}S_m(\mathbf k)+k^2\mathsf P^l{}_{i}S_l(\mathbf k)\Big)
=
\mathsf P^m{}_{i}S_m(\mathbf k)
=
S_i(\mathbf k),
\end{aligned}
\end{equation}
where the mixed terms vanish because $\mathsf P^m{}_{i}k_m=0$, and the last step uses the transversality condition $k^mS_m=0$.
Eq.~\eqref{eq:vector-source-extraction}  implies  \(\pi^{lk}{}_{i}(-i\mathbf k)\,\pi^{i\,mn}(-i\mathbf k')
\left\langle
S_{lk}(\mathbf k,\eta')\,S_{mn}(\mathbf k',\eta'')
\right\rangle
=
\left\langle
S(\mathbf k,\eta')\,S(\mathbf k',\eta'')
\right\rangle
e_i(\hat{\mathbf k})\,e^i(\hat{\mathbf k}')
\), connecting Eq.~\eqref{noise1} and Eq.~\eqref{eq:omegaomega-projected}.

For a single polarization, the remaining shear contraction gives
\begin{equation}
\pi^{lk}_{(\lambda)}\,\pi^{mn}_{(\lambda)}
\left(\delta_{lm}\delta_{kn}+\delta_{ln}\delta_{km}\right)
=
\frac{1}{k^2},
\end{equation}
while the bulk/trace piece proportional to $\delta_{lk}\delta_{mn}$ drops out because $\delta_{lk}\pi^{lk}_{(\lambda)}=0$. Hence the per-polarization kernel is
\begin{equation}
\mathcal N_S^{(\lambda)}(k,\eta)
=
2\mathcal T_0\,\frac{1}{k^2}\,\zeta_s
=
\frac{2\alpha \mathcal T_0^4}{k^2}.
\end{equation}
Summing the two polarizations then gives the total kernel
\begin{equation}
\mathcal N_S(k,\eta)
=
\sum_{\lambda=1}^{2}\mathcal N_S^{(\lambda)}(k,\eta)
=
2\mathcal T_0\,\frac{C_V}{k^2}\,\zeta_s
=
\frac{4\alpha \mathcal T_0^4}{k^2},
\end{equation}
with
\begin{equation}
C_V=2.
\end{equation}
Here $C_V=\operatorname{tr}(\mathsf P_{ij})=2$ is simply the value of the summed transverse trace in the compact total-kernel notation.
Keeping the projector factor $C_V/k^2$ explicit is essential: dropping the $k^{-2}$ makes the final $\mathcal P_{F'}$ dimensionful instead of dimensionless.

\section{Late-time asymptotic regimes}
\label{app:Late-Time Asymptotic}

Starting from Eq.~\eqref{eq:exact-spectrum},
\begin{equation}
\mathcal P_{F'}=
\frac{8\kappa^2\alpha \mathcal T_0^4 k}{\pi^2}
\; y^{4\beta}\,
e^{-2d_1 y^{1-\beta}+2d_2 y^{\beta+1}}
\int_y^{y_0} dx\; x^{-4\beta}
e^{2d_1 x^{1-\beta}-2d_2 x^{\beta+1}},
\qquad y=-\eta>0,\label{eq:appendix-exact-spectrum}
\end{equation}
we consider the late-time limit $y\to 0^+$, equivalently $a\to\infty$.
Given the slow-roll evolution of the scale factor,
\begin{equation}
a(\eta)=A_0(-\eta)^{-\beta},
\end{equation}
and define
\begin{equation}
f(x):=x^{-4\beta}e^{2d_1 x^{1-\beta}-2d_2 x^{\beta+1}},
\qquad
I(y):=\int_y^{y_0} f(x)\,dx.
\end{equation}
Then
\begin{equation}
\mathcal P_{F'}=
\frac{8\kappa^2\alpha \mathcal T_0^4 k}{\pi^2}\,
\frac{I(y)}{f(y)}.
\end{equation}
For $\beta>1$, one has $f(y)\to\infty$ as $y\to 0^+$, and the integral $I(y)$ also diverges, so the ratio is of $\infty/\infty$ type. We can then apply L'H\^opital's rule to obtain
\begin{equation}
\frac{I(y)}{f(y)}
\;\longrightarrow\;
\frac{I'(y)}{f'(y)}
=
\frac{-f(y)}{f'(y)}
=
-\frac{1}{(\ln f)'(y)}.
\end{equation}
Now
\begin{equation}
\ln f(y)=-4\beta\ln y+2d_1 y^{1-\beta}-2d_2 y^{\beta+1},
\end{equation}
so
\begin{equation}
(\ln f)'(y)
=
-\frac{4\beta}{y}
+2d_1(1-\beta)y^{-\beta}
-2d_2(\beta+1)y^{\beta}.\label{eq:appendix-log-derivative}
\end{equation}
Writing
\begin{equation}
d_1=\frac{c_1}{\beta-1},
\qquad
d_2=\frac{c_2}{\beta+1},
\end{equation}
Eq.~\eqref{eq:appendix-log-derivative} becomes
\begin{equation}
(\ln f)'(y)
=
-\frac{4\beta}{y}
-2c_1 y^{-\beta}
-2c_2 y^{\beta}.
\end{equation}
In the slow-roll background assumed here, $1<\beta<2$, so $y^{\beta}\to 0$ as $y\to 0^+$. Since $y^{-\beta}$ diverges faster than $y^{-1}$, the dominant term is clearly $-2c_1 y^{-\beta}$, hence
\begin{equation}
\frac{I(y)}{f(y)}
\simeq
\frac{y^\beta}{2c_1}.
\end{equation}
Therefore
\begin{equation}
\mathcal P_{F'}
\simeq
\frac{8\kappa^2\alpha \mathcal T_0^4 k}{\pi^2}\cdot \frac{y^\beta}{2c_1}.
\end{equation}
Substituting $c_1=2\kappa\alpha \mathcal T_0^3 A_0$ and $y=-\eta$ gives
\begin{equation}
\mathcal P_{F'}
=
\frac{2\kappa \mathcal T_0 k}{\pi^2 A_0}(-\eta)^\beta.
\end{equation}
Finally, recalling
\begin{equation}
a(\eta)=A_0(-\eta)^{-\beta},
\qquad
\frac{(-\eta)^\beta}{A_0}=\frac{1}{a(\eta)},
\end{equation}
we obtain Eq.~\eqref{eq:strict-asymptotic}:
\begin{equation}
\mathcal P_{F'}(k,\eta)\simeq
\frac{2\kappa \mathcal T_0 k}{\pi^2 a(\eta)}.
\end{equation}
The factor $\alpha$ in the shear viscosity cancels in the strict late-time limit because it appears both in the noise amplitude and in the dissipative coefficient $c_1$.

Eq.~\eqref{eq:strict-asymptotic} is the strict asymptotic as $y\to0^+$ at fixed $\alpha>0$ and $\epsilon>0$. For small $r_{\zeta,*}$ and small $\epsilon$, however, the onset of that regime is extremely delayed. The ratio of the local viscous term to the Hubble term is $r_\zeta(N)=r_{\zeta,*}e^{\epsilon N}$ as given in the main text.  A useful indicator is the crossover between the Hubble term and the viscous term in
\begin{equation}
-(\ln f)'(y)=\frac{4\beta}{y}+2c_1 y^{-\beta}+2c_2 y^{\beta}.
\end{equation}
Balancing the first two terms gives
\begin{equation}
\frac{4\beta}{y}\sim 2c_1 y^{-\beta}
\qquad\Longrightarrow\qquad
y_*\sim \left(\frac{c_1}{2\beta}\right)^{1/(\beta-1)}.
\end{equation}
Since $\beta-1=\epsilon/(1-\epsilon)\approx\epsilon$, the turnover happens around
\begin{equation}
r_{\zeta,*}e^{\epsilon N_*}\sim1,
\end{equation}
which is extremely delayed when $r_{\zeta,*}\ll1$ and $\epsilon$ is small.

When $y\gg y_*$, the exact spectrum remains in a long pre-asymptotic regime. In this regime the numerically accurate late-time approximation is
\begin{equation}
\mathcal P_{F'}
\approx
\frac{8\kappa^2\alpha \mathcal T_0^4 k}{\pi^2(4\beta-1)}(-\eta),
\end{equation}
namely Eq.~\eqref{eq:pre-asymptotic}, whereas the strict asymptotic behaviour becomes a better approximation much deeper into the $y\to0^+$ limit. It is therefore best to keep Eq.~\eqref{eq:strict-asymptotic} and Eq.~\eqref{eq:pre-asymptotic} as two different asymptotic formulas with different domains of applicability:
\begin{equation}
\mathcal P_{\text{Eq.~\eqref{eq:strict-asymptotic}}}(k,\eta):=
\frac{2\kappa \mathcal T_0 k}{\pi^2 a(\eta)},
\qquad
\mathcal P_{\text{Eq.~\eqref{eq:pre-asymptotic}}}(k,\eta):=
\frac{8\kappa^2\alpha \mathcal T_0^4 k}{\pi^2(4\beta-1)}(-\eta).
\end{equation}
A practical turnover scale can be obtained by equating the two asymptotic approximations,
\begin{equation}
\mathcal P_{\text{Eq.~\eqref{eq:strict-asymptotic}}}(z_{\rm turn})=\mathcal P_{\text{Eq.~\eqref{eq:pre-asymptotic}}}(z_{\rm turn}),
\qquad
z:=|k\eta|=ky,
\end{equation}
which gives
\begin{equation}
z_{\rm turn}
=
k\left(\frac{4\kappa\alpha \mathcal T_0^3 A_0}{4\beta-1}\right)^{1/(\beta-1)}.
\end{equation}
Near $z\sim z_{\rm turn}$ neither pure asymptotic is uniformly more accurate, and one should revert to the exact expression Eq.~\eqref{eq:appendix-exact-spectrum}. The late time power spectrum also evolves very quickly to become independent from the choice of initialization time. Denote $C=\frac{8\kappa^2\alpha\mathcal T_0^4k}{\pi^2}$ and take the physically accessible pre-asymptotic regime initialized at time $y_0$ as an example, 
\[
\mathcal P(y;y_0)\simeq
\frac{Cy}{4\beta-1}
\left[1-\left(\frac{y}{y_0}\right)^{4\beta-1}\right].
\]
Noting $\left(\frac{y}{y_0}\right)^{4\beta-1}
=e^{-(3+\epsilon)\Delta N}$, it is evident that the contribution from $y_0$ vanishes within the first few e-folds and the power spectrum approaches its pre-asymptotic behaviour very fast.

\section*{Acknowledgements}
Z.Q. thanks Arko Bhaumik for useful comments.
S.B. is supported in part by a start-up grant from the Indian Statistical Institute and by an Honorary Fellowship at the University of Edinburgh.
A.B. acknowledges partial support from STFC.
R.O.R. is partially supported by research grants from Conselho Nacional de Desenvolvimento Cient\'{\i}fico e Tecnol\'ogico (CNPq), Grants No.~319307/2025-5 and from Funda\c{c}\~ao Carlos Chagas Filho de Amparo \`a Pesquisa do Estado do Rio de Janeiro (FAPERJ), Grant No. E-26/200.415/2026.

\end{document}